\documentclass[a4paper,fleqn,useAMS,usenatbib]{mnras}

\pdfoutput=1

\usepackage{natbib}
\usepackage{mathptmx}

\usepackage[T1]{fontenc}
\usepackage{ae,aecompl}

\usepackage[dvips]{graphicx}
\usepackage{amsmath}
\usepackage{amsfonts}
\usepackage{amssymb}
\usepackage{comment}
\usepackage{bm} 
\usepackage[dvipsnames]{xcolor}
\usepackage[]{hyperref}
\hypersetup{
	colorlinks=true,	
	urlcolor=MidnightBlue,		
	hypertexnames=true,
	plainpages=false,
	naturalnames=true,
	pdftitle={Diving orbits galactic center},    
	pdfauthor={Zephyr Penoyre},     
	linkcolor=WildStrawberry,          
	citecolor=Periwinkle,        
}

\usepackage[percent]{overpic}
\usepackage{subfig}
\usepackage{csvsimple}
\usepackage{lscape}
\usepackage[shortlabels]{enumitem}

\DeclareMathAlphabet{\mathcal}{OMS}{ntxsy}{m}{n}   
\SetMathAlphabet{\mathcal}{bold}{OMS}{ntxsy}{b}{n} 

\makeatletter
\def\env@matrix{\hskip -\arraycolsep 
  \let\@ifnextchar\new@ifnextchar
  \array{*{\c@MaxMatrixCols}c}}
\makeatother

\title[Reevaluating the loss-cone formalism]{Chaos and loss: reevaluating the loss-cone formalism in the context of chaotic diving orbits in flattened potentials}
\author[z. penoyre et al.]{zephyr penoyre$^{1}$\thanks{E-mail:
\href{mailto:zephyrpenoyre@gmail.com}{zephyrpenoyre@gmail.com}}, Elena Maria Rossi$^{1}$, Nicholas C. Stone$^{2,3}$
\\
$^{1}$Leiden Observatory, Leiden University, P.O. Box 9513, 2300 RA Leiden, the Netherlands\\
$^{2}$Department of Astronomy, University of Wisconsin, Madison, WI, 53706, USA\\
$^{3}$Racah Institute of Physics, The Hebrew University, 91904, Jerusalem, Israel
}

\date{Submitted 29/10/2026}

\pubyear{2025}

\begin{document}
\label{firstpage}
\pagerange{\pageref{firstpage}--\pageref{lastpage}}
\maketitle

\begin{abstract}
The region close to a massive black hole is one of the few truly spherical geometries in galactic dynamics. Stellar structures, which dominate from pc to kpc scales, are generally at least flattened (axisymmetric) if not more irregular and time evolving. Here we show that orbits that pass between these diverse geometries are commonly chaotic, especially those which \textit{dive} - visiting their minimum possible periapse. We show that these chaotic diving orbits are common across a wide range of potentials, and can lead to large variation in angular momentum orbit to orbit. We characterise the transition from regular to chaotic orbits with a notional critical angular momentum, $j_d$. We then model the flux of orbits into the loss cone including the effects of these diving orbits combined with evolution through two-body scattering. Systems which cross into the chaotic region are quickly advected away, and thus there is a large net flux across the boundary. We formulate the problem as a probability density, rather than a number density, allowing us to find a true equilibrium state. We find that the rate of disruptions when including diving orbits is boosted by a multiplicative factor of $1+\ln(j_d/j_t)$, where $j_t$ is the angular momentum corresponding to a disruption. This boost can increase disruption rates by an order of magnitude compared to the classical loss cone formalism.
\end{abstract}


\section{Introduction}
Nuclear star clusters (NSCs) are the densest stellar systems in the Universe, with central stellar densities reaching up to millions of solar masses per cubic parsec \citep{Neumayer20}. Most NSCs host a central massive black hole (MBH), whose gravitational field acts as a sink and disruptor for stars, binaries, and compact objects that evolve onto orbits of very low angular momentum. The resulting close encounters produce a wide range of observable phenomena, including tidal disruption events (TDEs; \citealt{Hills75}), hypervelocity stars (HVSs; \citealt{Hills88}), and extreme mass-ratio inspirals (EMRIs; \citealt{Hils95}), making galactic nuclei unique laboratories for studying stellar dynamics, black hole demographics, and strong-field gravity. Current and forthcoming electromagnetic and gravitational-wave facilities promise  increasingly precise measurements of these phenomena, motivating a deeper theoretical understanding of the mechanisms that drive objects towards the MBH.

The long-standing framework for describing this process is the {\it loss-cone theory}, which characterizes how objects are supplied to the lowest angular momentum regions.  
In the classical picture, the loss cone is populated primarily through {\it collisional relaxation}: repeated weak gravitational encounters with the surrounding stellar population gradually diffuse objects in angular momentum until they reach the critical value required for disruption or other loss processes (\citealt{Frank76, Lightman77, Cohn78}; see \citealt{Merritt13}; \citealt{Stone20} for recent reviews). This mechanism has formed the basis of most theoretical estimates of tidal disruption rates \citep{Syer99, Wang04, Stone16, Pfister20, Chang25, Hannah25}, binary separation rates \citep{Perets07}, and EMRI rates \citep{Sigurdsson97, Hopman05, Raveh21, Broggi22, Qunbar24, Mancieri25} over the past several decades.

However, real galactic nuclei are not perfectly spherical. Observations of the Milky Way and many nearby galaxies reveal significant departures from spherical symmetry, particularly in the form of flattened nuclear stellar clusters \citep{Georgiev14} and nuclear stellar discs \citep{Sormani22}. In aspherical systems, the global potential applies torques, allowing stars to evolve {\it collisionlessly} onto low angular momentum trajectories. Previous studies have shown that axisymmetric \citep{Magorrian99, Vasiliev13, Kaur24} and triaxial \citep{Merritt04, Vasiliev15} potentials support families of centrophilic orbits that can enhance the supply of objects to the loss cone relative to the classical collisional channel.  Orbits in these aspherical potentials become vulnerable to disruption during periods of low angular momentum, when they may {\it dive} to very small periapse.  Past examinations of axisymmetric potentials generally found relatively modest enhancements, a factor of a few, and usually interpreted the responsible orbits as regular families that precess through phase space (though see \citealt{Magorrian99}).

In \citet{Penoyre25}, hereafter PRS25, we revisited this problem using a realistic model of the Milky Way potential extending to $\sim 200$ pc \citep{Sormani22}. We showed that stars traversing the transition between the nearly spherical inner regions and the significantly flattened nuclear stellar cluster and nuclear stellar disc do not predominantly occupy regular centrophilic orbits. Instead, they possess only two integrals of motion (energy $E$ and $z$-component of angular momentum $\ell_z$), and can hence be termed {\it chaotic diving orbits}. These chaotic trajectories undergo rapid excursions in total angular momentum, and for sufficiently small conserved $\ell_z$, these excursions allow these stellar objects to reach the loss cone even absent any two-body relaxation. PRS25 further suggested that this collisionless mechanism could dominate the supply of disruption events (for binaries and single stars) and potentially enhance their rates by more than an order of magnitude.

A limitation of PRS25, however, was that the collisional and collisionless channels were treated separately. The estimate of the enhancement due to diving orbits neglected the depletion of the diving region and did not self-consistently account for the interplay between chaotic angular-momentum evolution and classical two-body diffusion. Consequently, while the existence of chaotic diving orbits was established, their quantitative contribution to the steady-state loss-cone flux remained uncertain.

In this paper we develop a unified framework that simultaneously incorporates both mechanisms. We investigate a suite of simplified axisymmetric potentials spanning a broad range of flattenings and structural parameters, allowing us to isolate the essential physics governing chaotic diving orbits. We first demonstrate that chaotic diving orbits are a generic feature of even moderately flattened axisymmetric potentials, and they disappear well within the sphere of influence of the black hole.
We then combine chaotic angular-momentum advection with classical diffusion to derive equilibrium solutions for the occupation probability of phase space and compute the corresponding disruption rates. Our results  demonstrate that collisionless angular-momentum evolution in axisymmetric galactic nuclei is not merely a correction to classical loss-cone theory, but can represent a dominant channel for feeding massive black holes.

In section 2 we introduce the criteria for an orbit to come close enough to a MBH for disruption, the mechanisms by which it can do so, and specifically a notional critical angular momentum, $j_d$, below which we expect orbits to be chaotic. In section 3 we use a broad range of model potentials to find heuristic values of $j_d$ as a function of the test potential, energy and z-angular momentum. In section 4 we build a self-consistent combined framework for the redistribution of systems, of a fixed energy, in angular momentum space incorporating the effects of local scattering and chaotic angular momentum redistribution. In section 5 we introduce a toy model that can simulate the evolution of the combined framework to an equilibrium state, and from this we test the effective boost to the rate and properties of disruptions.  We summarize and conclude in section 6.

\section{The loss cone and the loss wedge}
\label{sec:coord}

In an axisymmetric potential a trajectory\footnote{Following the terminology employed in PRS25 we use \textit{orbit} to refer to the motion over a single radial period, and \textit{trajectory} to refer to the long term motion over many orbits.} does not conserve total (specific) angular momentum, $\mathbf{l}$, only the z-component thereof, $l_z$, and (specific) energy, $\varepsilon$.

A disruption of any type requires a close passage with the central singularity of mass $m$, inside the regime where the local potential is dominated by the point mass. Thus the periapse distance, $r_{\rm p}$, is approximately\footnote{This formula is exact if the energy of the incoming system, relative to the MBH, is 0. More precisely the factor approximated here as 2 is $1+e(\varepsilon,l)$ where $e$ is the eccentricity. $e\rightarrow 1$ as $\varepsilon \rightarrow 0$ regardless of $l$. Almost all systems of interest in this work approach the MBH from large distances, and are well approximated as free-fall from infinity with $\varepsilon \approx 0$.}
\begin{equation}
\label{eq:rp_lp}
r_{\rm p} \approx \frac{l^2}{2Gm}
\end{equation}
where $l=|\mathbf{l}|$ is the instantaneous total angular momentum at periapse.

We can think of a particular system as having a \textit{tidal radius}, $r_{\rm t}$, the distance from the singularity within which the tidal forces overcome its own self gravity, and it will disrupt.\footnote{For a system with physical extent, $a$, and mass ratio of the MBH to the system, $Q$ ($\gg 1$), this can be found by Taylor expanding the tidal potential of the point mass to be approximately $r_{\rm t} \sim Q^{1/3}a$. This applies equally well for a single star (where $a$ would represent the stellar radius), a binary system (where $a$ would represent the semi-major axis) or a star cluster (where $a$ would represent the scale length of the cluster's potential, of order the half-mass radius).} We can re-express this, using equation \ref{eq:rp_lp}, as a critical \textit{tidal angular momentum},
\begin{equation}
l_{t}\approx \sqrt{2Gmr_t}.
\end{equation}
Any system with $l < l_{t}$ at periapse is thus disrupted. This region of parameter space is often called the \textit{loss-cone} (a cylinder in angular momentum space alone, but accounting for variations in energy it becomes a conical surface of section).

For an oblate (disk-like) axisymmetric potential, trajectories of a given energy have a maximum possible angular momentum, $l_c(\varepsilon)$, corresponding to a closed circular trajectory in the equatorial plane (or any plane for a spherical potential). It is useful to work in rescaled and normalised transformations of any angular momenta, $l_i$,
\begin{equation}
j_i=\left| \frac{l_i}{l_c}\right|
\end{equation}
which can only take values between 0 and 1.

As $l_c$ scales with the size of the orbit, and $l_t$ scales with the size of the object, we generally expect $j_t \ll 1$. The total angular momentum is the magnitude of a vector, and thus the volume of phase space corresponding to the loss cone is tiny:
\begin{equation}
V_C=V(j<j_t) \propto j_t^2 \ll 1.
\end{equation}
Therefore, disruptions are infrequent and have little effect on the overall equilibrium state of the galaxy, except at relatively small radii, where they can weakly deplete radial orbits \citep{Cohn78}. 

Previous exploration of the rates and character of such disruptions most often assumed spherical symmetry, and thus the evolution by perturbations of an otherwise conserved total angular momentum. These perturbations can take any form, but the most ubiquitous and commonly discussed are the random velocity kicks caused by two-body relaxation: hyperbolic encounters between the object of interest and other background objects. We will refer to this spherical, diffusive problem as the \textit{classical loss-cone} formalism. 

To understand the effect of scattering events it is useful to consider the dynamical evolution of one \textit{target} object interacting with a \textit{background} population. Any sufficiently close encounter with a background object will, almost instantaneously, cause a velocity kick. As encounters rely on transient spatial proximity, rather than globally similar orbits, the frequency and magnitude of these kicks is largely independent of the target object's current angular momentum. Assuming a non-rotating isotropic background the net effect of many scatterings should, to first order, cancel (there should be as many kicks in any one direction as its opposite). 

Thus the predominant effect of two body scatterings is dispersive, leading to slow diffusion through angular momentum space.  For the nearly radial orbits relevant for loss cone problems, perturbations in energy are subdominant to those in angular momentum \citep{Frank76}.  
Thus we can expect the total angular momentum of a target to change from orbit to orbit following $j\rightarrow j+\mathcal{J}(\sigma)$ where $\mathcal{J}(\sigma)$ is some random perturbation with mean 0 and variance $\sigma$. The quantity $\sigma$ represents the average over many perturbative encounters, of varying magnitude, over one whole orbit of the target\footnote{For direct comparison with the existing literature, $\sigma^2$ is equal to the product of the orbital period and the diffusion coefficient $\langle \left(\Delta j \right)^2 \rangle_t$, where the subscript $t$ denotes the orbit average. See sections \ref{sec:numberandrate} and \ref{sec:classical} for a full comparison with the notation used commonly in the literature and specifically in \citet{Merritt13}.}. Thus, in the classical formalism, only systems with $j$ within a few $\sigma$ of $j_t$ have significant chance of immediate disruption.

However, as we showed in PRS25, in a non-spherical potential, most systems with sufficiently low $j$ will be chaotic, experiencing large, random jumps in $j$ every orbit. As $j$ is always $\geq j_z$, a necessary condition for a system to disrupt in the absence of scattering is that $j_{z}<j_t$. Given long enough, the random chaotic precession of $j$ will bring a system into the loss cone. The larger 1D region of orbits with $j_{z}<j_t$ is often called the \textit{loss wedge} \citep{Magorrian99}.

As we will show in section \ref{sec:est_lambdad} these chaotic orbits vary in $j$ approximately uniformly. We can introduce a notional critical angular momentum, $j_d$,  beneath which we expect chaotic orbits.  We shall find approximate values of $j_{\rm d}$ for a wide range of potentials in section \ref{diving}. Within this region we expect the total angular momentum of a target to vary randomly between $j_z$ and $j_d$ from orbit to orbit. Thus any system with $j_z<j_t$ and $j<j_d$ has a chance of disruption from chaotic angular momentum redistribution alone. As a consequence, the volume of parameter space corresponding to the loss wedge is substantially larger than the loss cone:
\begin{equation}
V_W=V(j_z<j_t, j<j_d) \propto j_t j_d.
\end{equation}

As long as $j_d>j_t$ the loss cone is a subset of the loss wedge, and thus we should expect a boost to the rate of disruptions. However, diving orbits cannot on their own affect systems with $j>j_d$, or change $j_z$.  
While the loss wedge of an axisymmetric nucleus will produce disruptions at a (high) drain-limited rate at early times, it must ultimately transition to disrupting stars at a (lower) replenishment-limited rate, with an empty or partially empty loss wedge slowly being refilled by two-body scattering \citep{Magorrian99}.  We are interested in finding stationary states where the combined rate of depletion by disruptions is matched by the rate of refilling via scattering. A full picture of the rate and type of disruptions must thus include both the effects of chaotic precession of $j$ and scattering in $j$ and $j_z$.


In PRS25 we only considered scatterings only as a mechanism of refilling the loss wedge through the $j_z= j_t$ barrier, ignoring the effect of depletion outside of this barrier. In simple solutions to the classical formalism the number density (and hence disruption rate) of systems near the loss cone boundary can be reduced by a significant factor of $2|\ln j_t|$. We should expect similar behaviour at the loss wedge boundaries, and thus expect the rates we previously presented to be an overestimate.

In this paper we will present a \textit{combined} formalism, establishing equilibrium distributions and disruptions rates corresponding to a given orbital energy. In our simplified model the equilibrium state can be parametrised entirely in terms of $j_t$, $j_d$ and $\sigma$.

\section{Diving orbits}
\label{diving}
For a given combination of $\varepsilon$ and $j_z$ in an axisymmetric potential, there will be some regular trajectories, 
which evolve predictably and quasi-periodically.  For sufficiently low $j_z$ values, there will also be chaotic trajectories, which diverge exponentially from their their past states over time.

In PRS25 we examined a realistic axisymmetric model of the MW GC. The trajectories that reach the minimum possible periapse, following equation \ref{eq:rp_lp}, are those which approach $j \rightarrow j_z$ at periapse. We refer to any trajectory that achieves this as \textit{diving}, and showed that in general axisymmetric potentials such diving trajectories are typically chaotic (this is in contrast to nearly Keplerian potentials, where regular trajectories may librate into a diving, $j \approx j_z$, state; \citealt{Vasiliev13}).

The chaotic region of parameter space is complex, often with regular substructure perforating it. Whilst the boundary of the chaotic region, at high $j$, cannot be expressed simply, the intention of this section is to formulate and test a characteristic $j_d$, an angular momentum which can be approximated as the transition from regular to chaotic diving orbits, for a range of potentials and integrals of motion $\varepsilon$ and $j_z$.  In reality, no such single angular momentum exists, but estimating $j_d$ is both a useful proxy for the extent of the diving region, and allows powerful mathematical simplifications of the diffusion processes we detail in later sections. 

\subsection{Model potential}

We will work with test potentials of the form of a Keplerian point mass black hole, of mass $m$, and an extended mass distribution in the form of a Miyamato-Nagai (MN) disk \citep{Miyamoto75}. This profile is chosen because it is an axisymmetric model with finite mass, $M$, a well defined potential and density, and can be smoothly transformed to a spherical potential depending on the parameters.

The overall mass and length scale of the problem (and thus the associated timescales) can be factored out, though for clarity here we'll define them. For a MN disk with scale length, $b$, we can write the full potential as
\begin{equation}
\label{eq:test_potential}
\Phi = -\frac{GM}{b}\left(\mu \tilde{r}^{-1} +  \left[\tilde{R}^2 + \left(\alpha  +\sqrt{1+\tilde{z}^2}\right)^2\right]^{-\frac{1}{2}}\right),
\end{equation}
where $\tilde{r}=r/b$ is a dimensionless spherical radius, and cylindrical coordinates $\tilde{R}$ and $\tilde{z}$ have likewise been non-dimensionalized. The first term on the RHS comes from the MBH, the second comes from the MN disk, and $\mu=m/M$ is their \textit{mass ratio}. 

The dimensionless constant $\alpha$ defines the degree of \textit{asphericity} of the MN disk. Isopotential contours of the MN disk are those where
\begin{equation}
1+\alpha^2 + \tilde{R}^2 + \tilde{z}^2 +2\alpha\sqrt{1+ \tilde{z}^2} =\mathrm{constant}.
\end{equation}
For $\tilde{z}\ll 1$ 
\begin{equation}
(1+\alpha)^2 + \tilde{R}^2 + (1+\alpha)\tilde{z}^2 +O(\tilde{z}^4) =\mathrm{constant},
\end{equation}
and the MN disk thus fits the form for a flattened Plummer model \citep{Evans94} with a spheroidal radius 
\begin{equation}
r' = \sqrt{r_s^2 + R^2 + \left(\frac{z}{q}\right)^2},
\end{equation}
with scale radius $r_s = (1+\alpha)b$ and a flattening parameter (i.e. the ratio of minimal to maximal distances to the isopotential contour) $q = (1+\alpha)^{-\frac{1}{2}}$. At intermediate $\tilde{z}$ the degree of flattening increases, and the potential is disk-like rather than spheroidal. For $\tilde{z} \gg 1$ the potential tends towards a spherical Plummer model with the same scale radius, and at large distances ($\tilde{r}\gg \alpha$) resembles a Keplerian point mass.


It is useful to define an influence radius, $r_0$, within which the MBH dominates the potential. There are a variety of possible ways to define this quantity\footnote{For example, the radius at which the mass enclosed from the extended distribution is equal to that of the MBH \citep{Merritt13book}.}.  
For our purposes, we will define it as
\begin{equation}
\partial_R \Phi_{MBH}|_{r_{0},0} = \partial_R \Phi_{Ext}|_{r_0,0},
\end{equation}
which is equivalent to saying that $r_0$ is the radius of an orbit in the disk ($z=0$) where both components contribute equally to the circular velocity (or the radius at which orbits would have the same period if the two potentials were considered in isolation). This gives
\begin{equation}
r_0=\frac{1+\alpha}{\sqrt{\mu^{-2/3} - 1}}b,
\end{equation}
which goes to infinity as $\mu \rightarrow 1$, and becomes ill-defined for $\mu>1$.

\subsection{Trajectories}

\begin{figure*}
\includegraphics[width=\textwidth]{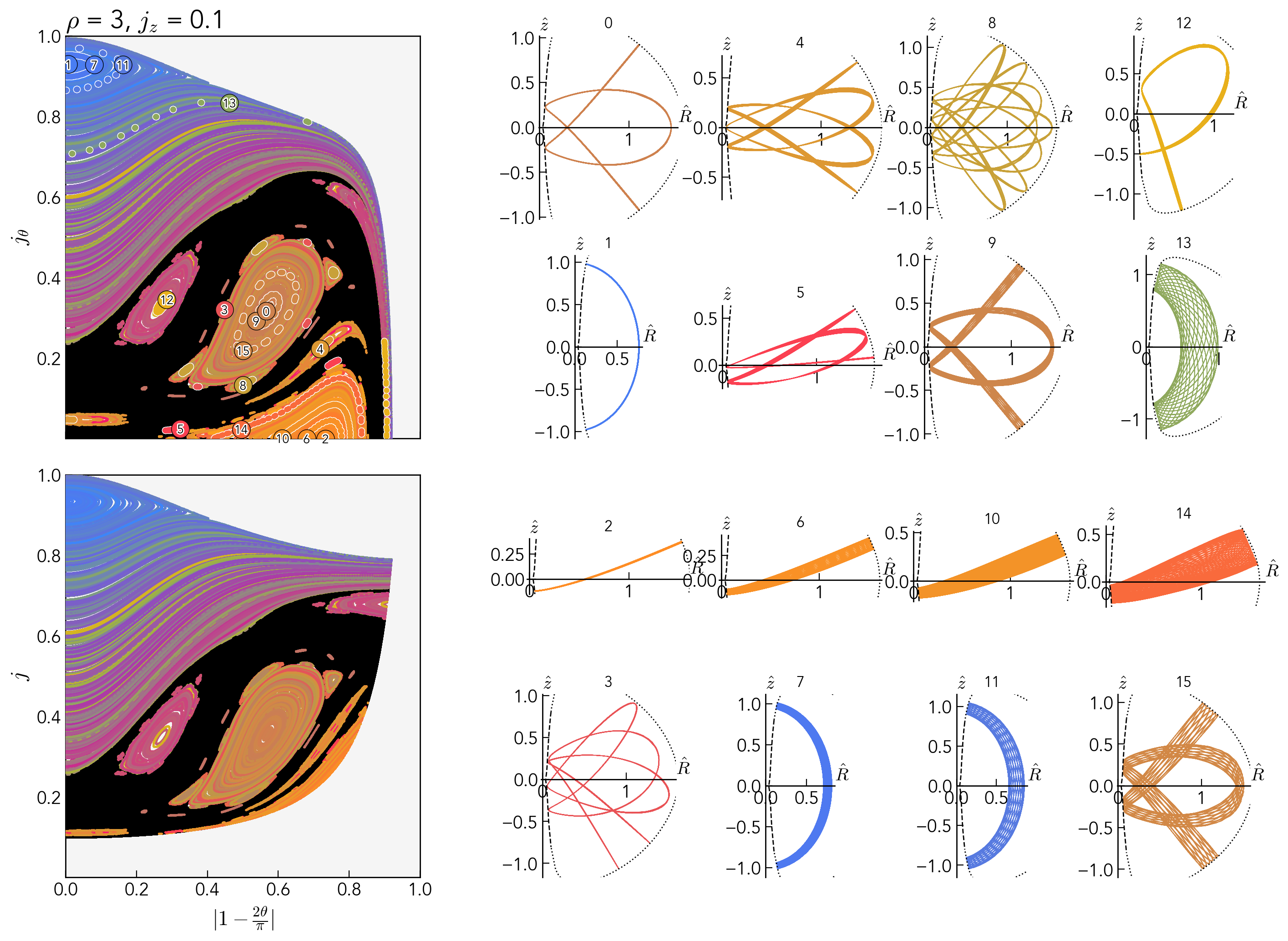}
    \caption{The full set of trajectories in a MN disk plus MBH potential, with mass ratio $\mu=1/30$ and asphericity $\alpha=1$.  Trajectories are shown for dimensionless circular orbit radius $\rho = R_{\rm c}/r_0= 3$ and $j_z=0.1$; the influence radius is $r_0=0.68 b$. On the left we show surfaces of section (SoSs) depicting the state at periapse in terms of $\theta$ and $j_{\theta}$ (top) and $j$ (bottom). We sample 2000 random initial states and follow the trajectories for 2000 orbits each. Chaotic trajectories are coloured black, and regular trajectories are given a colour based on their initial state. To the right we plot examples of 16 regular trajectories in cylindrical coordinates (in units of $R_c$). The boundary of the explorable region by any trajectory of this energy and angular momentum is shown with a dotted line. We choose these trajectories by selecting those that take up minimal area in the SoS, which highlights closed orbits (including harmonics) and some-near closed orbits. We summarise some briefly here: tube (closed - 1, regular - 7, harmonic - 11, 13), banana (closed - 2, regular - 6, 10, harmonic - 14), fish (closed - 0, regular - 9, harmonic - 15), pretzel (closed - 12). More complex harmonic orbits exist which are period doublings of other orbit families, such as trajectories 3, 5 and 8. Refer to PRS25 for further details.}
    \label{fig:map}
\end{figure*}

Much as we can define a maximum angular momentum for a trajectory of a given energy, $l_c(\varepsilon)$, there is also a characteristic radius, $R_c(\varepsilon)$, and period, $T_c(\varepsilon)$, which characterise the typical orbital length and timescale (formulae for which can be found in section 2.2 of PRS25).  For oblate axisymmetric potentials, these variables are properties of the circular equatorial orbit at a given $\epsilon$.

The radius $R_c$ sets the physical length scale of the trajectory. The radii at apoapse, $r_a$, can vary orbit to orbit, but is constrained within $R_c\leq r_a \leq 2 R_c$.
Periapses may occur at much smaller radii, depending on the angular momentum (see equation \ref{eq:rp_lp}). We expect a transition depending on whether orbits stay confined to one dominant potential, or traverse between regions dominated by different potentials. Thus it is intuitive to reparametrize $\epsilon$ as a dimensionless circular equatorial radius,
\begin{equation}
\rho = R_c(\varepsilon)/r_0.
\end{equation}
A trajectory in a fixed axisymmetric potential has at most 4 degrees of freedom (DoF), as its 6D position and velocity state is constrained by 2 constant integrals of motion, $\varepsilon$ and $l_z$. We can treat one of these DoF as a phase-like coordinate, that denotes where, along the trajectory, the system currently is. The other 3 DoF determine the shape of the trajectory. We are free to choose any combination of independent DoF, but a useful convention (as used in PRS25) is to examine the spherical polar angle, $\theta$, and its corresponding canonical momentum\footnote{In PRS 25 we denoted $p_\theta$ as $l_\theta$, a confusing convention that we wish to correct here. Commonly $p_q$ is the generalized momentum associated with the variable $q$, and is equal to $\partial_{\dot{q}}\mathcal{L}$, where $\mathcal{L}$ is the Lagrangian of the system. Meanwhile $l_q$ usually denotes the component of the angular momentum parallel to the unit vector $\mathbf{\hat{q}}$. Thus in spherical coordinates $p_\phi=l_z$ and $p_\theta=l_\phi$. We will use $p_\theta$ here to emphasise the canonical coordinate momentum pair with $\theta$, though it may help with spatial intuition to remember that it's equivalent to the component of angular momentum in the $\bm{\hat{\phi}}$ direction.}, $p_\theta=r^2 \dot{\theta}$, which we can similarly express as a relative angular momentum,
\begin{equation}
j_\theta =\left|\frac{p_\theta}{l_c}\right|.
\end{equation}

The total (relative) angular momentum can thus be written as
\begin{equation}
\label{eq:j_simp}
j = \sqrt{j_\theta^2 + \frac{j_z^2}{\sin^2\theta}},
\end{equation}
which can approach its minimum possible value when $j_\theta \rightarrow 0$ (moving radially) and $\theta \rightarrow \frac{\pi}{2}$ (approaching the equatorial plane).

We can choose the final DoF to be $\dot{r}$, the radial velocity. We then reduce the 3D shape to a 2D projection which can be visualised and analysed, by examining the trajectory only at periapse ($\dot{r}=0$, $\ddot{r}>0$). This flattens the full behaviour onto a representative Surface of Section (SoS), as we show in figure \ref{fig:map}. All possible trajectories for a given potential, $\varepsilon$, and $l_z$ are represented. Closed regular trajectories return to the same locus of points, whilst other regular trajectories precess around these closed trajectories. In Figure \ref{fig:map}, we see a few of the major regular orbit families (e.g. tube, banana, fish and pretzel), some higher harmonics of these, and a large region spanned by chaotic diving orbits, coloured in black. Regular trajectories move predictably through a sequence of periapses, whilst for chaotic trajectories the state of the next periapse is essentially random.

Previous authors \citep{ME89,Fulton01} have noted that many of the regular trajectory families that we observe correspond to Lissajous curves, a family of trajectories defined by integer ratios of oscillation frequencies in two dimensions (here $z$ and an $R$-like coordinate that is capable of taking negative values). However, we note that we also recover trajectories outside of this family, including the higher order definite-rotation trajectories detailed in PRS25: these are extensions of the tube trajectory, which unlike many of the reversible-rotation trajectories captured by Lissajous curves, never double back on themselves. We also see here a number of period-doublings (new closed trajectories which resemble their parent but require twice as many oscillations to return to their original state) of commonly identified orbits such as the banana (panel 5 in figure \ref{fig:map}). We recover two families of period doubling of the fish orbit, one with reversible-rotation (panel 3) and one with definite-rotation (panel 8). Each of these period doublings appears just beyond the region occupied by their parent regular trajectory. Examining the SoS we can see regions of regular trajectories (not highlighted by our selection algorithm) that are similarly arrayed around the parent orbits that likely correspond to higher order period-doublings. Such period-doublings are a hallmark of a transition to chaos, though unlike many well studied chaotic systems (e.g. \citealt{May76}) our 2D space allows different families of period-doublings (definite or reversible rotation) and thus cannot be defined by a single control parameter.


\begin{figure*}
\includegraphics[width=\textwidth]{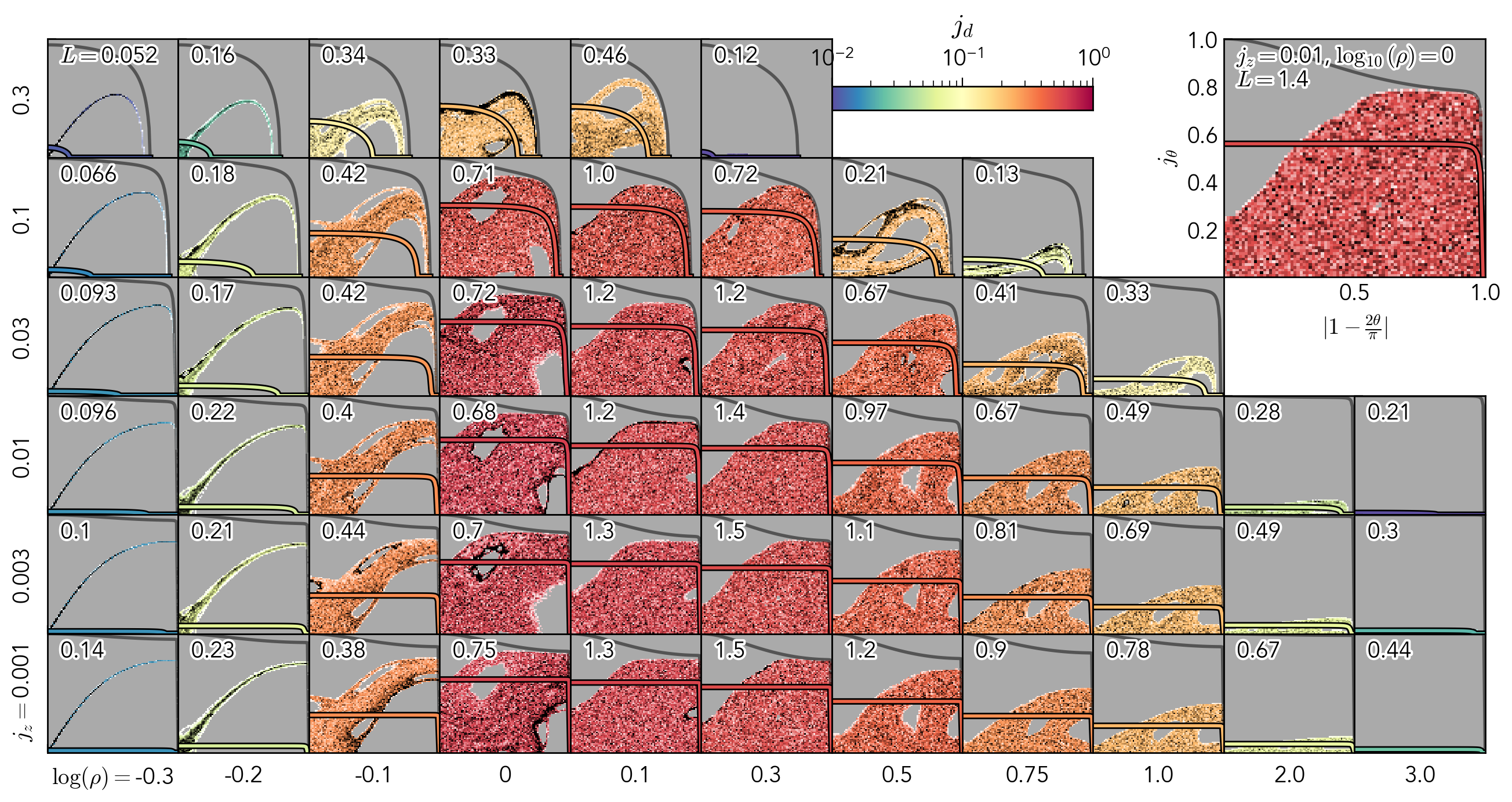}
    \caption{The shape and extent of the chaotic sea in our test potential (with $\mu=1/30$ and $\alpha=1$) for a variety of energies ($\rho$, columns) and z-angular momenta ($j_z$, rows).
    Each subplot shows the SoS at periapse, in the space of $j_\theta=|p_{\theta}/l_c|$ and $\theta$, both normalised to span from 0 to 1. In the top right we zoom in on the $j_z=0.01, \ \log(\rho)=0$ (corresponding to the circular radius being equal to the sphere of influence) panel; the axis labels and ranges in the zoom-in are the same as those used throughout the figure. Each panel shows a single trajectory with initial $\theta=\pi/2$ and $j_\theta=10^{-6}$ integrated over roughly 30,000 orbits. We plot a solid coloured line corresponding to a fixed $j=j_d$, such that $j\leq j_d$ occupies an equivalent area to that of the chaotic region. We also show the extent of the accessible region of any bound physical orbit in the SoS with a solid grey line. We show histograms of how frequently each region of the chaotic sea is visited by the trajectory, normalised by the median, ranging from white (approaching 0), to the colour corresponding to $j_d$ (1), to black ($\geq 2$). In the top left of each panel we note the Lyapunov exponent ($L$) measured for each trajectory, in units of $T_c$.}
    
    \label{fig:divingregions}
\end{figure*}

\subsection{Mapping out the diving region}

From figure \ref{fig:map}, we can see that neither the regular or chaotic orbits truly conserve $j$, as is expected for any axisymmetric potential. The chaotic sea is furthermore a geometrically complex region, and cannot easily be mapped to a rectangle-like zone.  Nevertheless, 
it is possible to estimate the area of the chaotic sea in the $\{j_\theta, \theta\}$ plane and calculate an effective $j_d$ from this.

In PRS25, we examined the full family of possible orbits, but here we use a more expedient method. We are specifically interested in diving trajectories, those which simultaneously approach $j_{\theta}\rightarrow 0$ and $\theta \rightarrow \frac{\pi}{2}$ (the bottom left corner of the $j_\theta$ SoS), which gives $j \rightarrow j_z$ and thus the minimum possible periapse. Thus to examine trajectories that may enter the loss cone we start with an initial periapse at $j_{\theta} \ll 1$ and $\theta=\frac{\pi}{2}$ (if we chose $j_{\theta}=0$ exactly the orbit would be confined to the equatorial plane, so hereafter we use $j_{\theta}=10^{-6}$).

This may seem like we are starting from our end goal, but given that our potential is time-independent any trajectory is time reversible, and thus integrating forward from this point is the same as tracing the orbit backwards from this point. 

We perform the integration in \texttt{AGAMA} \citep{Vasiliev18} using the same methodology as introduced in PRS25, and thus refer the reader to there for more details. We also performed comparative tests with the high-precision few-body integrator \texttt{SPACEHUB} \citep{Wang21} confirming that our trajectories are numerically robust.

Over many orbits we can examine if the trajectory is chaotic with \texttt{AGAMA}, by testing for chaotic behaviour. One way to quantify this is with a Lyapunov exponent, $L$, in units of $T_c(\varepsilon)$. A trajectory with $L\sim1$ diverges exponentially with time, for $L\ll1$ it takes $\sim\frac{1}{L}$ orbits to diverge meaningfully. If no exponential divergence is seen, it returns $L=0$, and we would consider the trajectory regular.

Whilst $L$ as measured by AGAMA is generally a robust and representative measure of chaos, we have discovered some edge cases where it breaks down. The most prominent of these occur for very radial orbits close to the MBH. These trajectories are regular, but small differences in phase are amplified to large differences in position and velocity near periapse, which can be erroneously attributed to exponential divergence. 

Thus we have formulated an improved test for the presence of chaotic behaviour, based on the dimensionality of the region each trajectory occupies in the SoS. Regular trajectories should be 1D, as they permit a third integral of motion that reduces the dimensionality of their parameter space \citep{Contopoulos63}. They may have occupied multiple regions of the SoS, corresponding to multiple different apo/periapses, as happens for example for orbit 7 in Figure \ref{fig:map} which has 4 distinct apoapsides, which appears as 2 regions in our SoS, when accounting for symmetries. As we approach a closed orbit the line-segments are smaller and smaller, eventually tending to 0D points. In contrast chaotic orbits do not obey a third isolating integral, and should span a 2D region of the SoS. Thus by measuring the dimensionality we can remove the false positives implied by the Lyapunov exponent calculation. We do this by calculating the correlation dimension of the region spanned by a given orbit, as is detailed in appendix \ref{ap:dim}.

Figure \ref{fig:divingregions} shows the extent of the chaotic sea corresponding to diving orbits, at various $\rho$ and $j_z$, in our test potential. Any two chaotic trajectories that overlap in this space are the same trajectory separated only by time, thus for a long enough integration (here we use 30,000 orbits) we see the full behaviour of this chaotic region. 

At greater distances (higher energies), the potential becomes approximately spherical and the chaotic region is small. The chaotic region is largest close to the influence radius of the MBH and then reduces again within the influence radius. Stable islands of regular orbits are sometimes present within the chaotic sea, and within the influence radius these can extend to $j_\theta=0$. This was not seen in PRS25, where we used a minimal smoothing radius for the MBH (which was there necessary for calculating the self consistent density-potential pair) which introduced another precession term at small radii.

The chaotic sea is generally very evenly sampled, as we would expect for ergodic chaos given that $j_{\theta}$ and $\theta$ are conjugate variables. In a minority of the panels we do see some regions that are consistently over/under dense. These tend to occur near the edge of regular regions, and suggest that trajectories can follow (or avoid) near regular patterns for many orbits before returning to the full chaotic sea. Numerical noise may also play a role here, perturbing orbits onto and out of regular regions (we might expect a similar behaviour from real astrophysical perturbations, which are not yet included here). 

At very low energies, deep within the sphere of influence, the chaotic region becomes a single arc with finite but diminishing width and a clear overdensity along its spine. This region is eventually too narrow for us to measure the dimensionality, but it is unclear if it is a consequence of numerical noise, or if it is a truly physical phenomenon, and how deep into the sphere of influence it could persist. Given that it represents a minimal area of the SoS it does not greatly affect the following analysis, but may represent an interesting physical phenomena, of chaos deep in the sphere of influence, that we leave for future work to explore.



\begin{figure*}
\includegraphics[width=\textwidth]{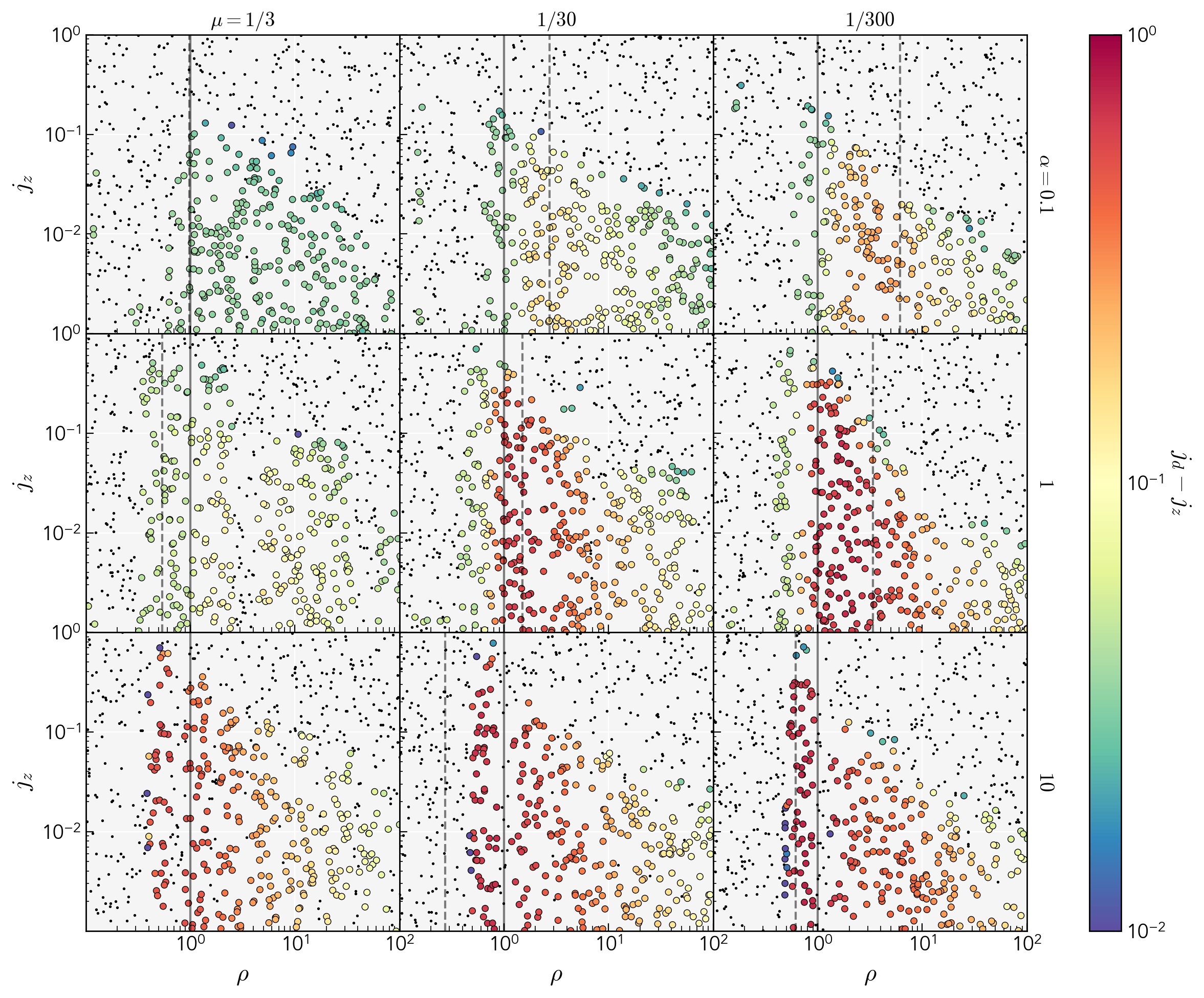}
    \caption{Estimations of $j_d-j_z$ over a range of potentials. Each panel shows a particular instance of our test potential (equation \ref{eq:test_potential}) with mass ratio, $\mu$, and flattening parameter $\alpha$. Each point is a trajectory with a particular energy and z-angular momentum (parametrised by $\rho$ and $j_z$). Trajectories are coloured by $\eta_{d}$, and trajectories that are not chaotic are shown as small black dots. The influence radius ($\rho=1$) is shown with a solid vertical line, and the scale radius of the potential with a dashed vertical line.}
    \label{fig:etad}
\end{figure*}

\subsection{Estimating $j_d$}
\label{sec:est_lambdad}

We wish to estimate a reasonable $j_d(\rho,j_z,\mu,q)$, i.e. a $j_d$ for any given potential, energy and z-angular momentum. We will do this by estimating the fraction of the physical region of $p_{\theta},\theta$ space spanned by a chaotic diving orbit, and then mapping this to an equivalent region of $j<j_d$.

A chaotic orbit that is ergodic will uniformly and randomly sample the available region in any space of any canonical coordinate pair. Thus a chaotic trajectory may visit any accessible state on any given orbit. The probability of any state is proportional, in these coordinates, to the area spanned by this state. 

In this section, we will use the coordinates of the unit-square SoS
\begin{equation}
\bar{x}=\left|1-\frac{2\theta}{\pi}\right| \ \ \mathrm{and} \ \ \bar{y}=j_\theta.
\end{equation}
Because they are a linear transformation of our canonical coordinates the uniform sampling is preserved, and because of their normalisation the probability of any state is directly proportional to the area, and directly equal to the area in the case where physical orbits can span the whole range of $\bar{x}$ and $\bar{y}$.

There is an arc in $\bar{x},\bar{y}$ space corresponding to a given $j$:
\begin{equation}
\bar{y}_j(\bar{x})=j\sqrt{1-\left(\frac{j_z}{j \kappa}\right)^2},
\end{equation}
where
\begin{equation}
\kappa = \sin \theta = \cos\frac{\pi \bar{x}}{2}.
\end{equation}
In the equatorial plane, $\bar{x}=0$, this goes to a maximum of $\bar{y}_j=j$. There is a critical polar angle, corresponding to $\bar{x}=\bar{x}_c$ where $\bar{y}_j \rightarrow 0$ satisfying $\kappa_c=\frac{j_z}{j}$.

To first order, for $j\rightarrow j+dj$, $\bar{y}_j \rightarrow \bar{y}_j+d\bar{y}_j$ where
\begin{equation}
d\bar{y}_j= j\frac{dj}{\bar{y}_j}
\end{equation}
and thus if this arc falls entirely within the chaotic region the probability of a momentary angular momentum between $j$ and $j + dj$ is
\begin{equation}
\begin{aligned}
p(j) dj \propto A(j+dj)-A(j) = \int_0^{\bar{x}_c} d\bar{y}_j d\bar{x} \\
=\frac{2 dj}{\pi} \int_{\kappa_c}^1 \kappa (1-\kappa)^{-\frac{1}{2}}(\kappa-\kappa_c)^{-\frac{1}{2}}d\kappa= dj
\end{aligned}
\end{equation}
where $A(j')$ is the area of the unit box with $j \leq j'$. This fact, that the probability of a given $j$ within the chaotic region is uniform, means that we can later approximate the effect of chaotic orbits as a uniform random resampling of $j_z<j<j_d$.

The total area of all orbits with $j$ less than or equal to some $j_d$ is 
\begin{equation}
A(j_d)=\int_{j_z}^{j_d}  dj=j_d-j_z.
\end{equation}
If we were to make the further assumption that \textit{all} orbits uniformly occupied some physical region of the unit box (with area $A_{\rm phys}$, as calculated in equation 27 of PRS25) then the probability that a random orbit of a given energy and angular momentum is a diving orbit is $P_{\rm dive}=A(j_d)/A_{\rm phys}$



To do this numerically, we discretize the $\bar{x},\bar{y}$ unit-square of the SoS into a grid and calculate the number of periapses that fall within each grid cell. We expect this number to be relatively uniform in the chaotic region, 0 outside of it, and for there to be some cells with low occupancy which span the edge of the region. We find the median occupancy of the cells (ignoring empty cells), and count the cells with occupancy greater than half of that median (a proxy for cells which are mostly or entirely in the chaotic region). We discard trajectories if they do not pass the dimensionality test, or that only span one cell.

Whilst it is not generally true that the chaotic sea spans the full arc in $\bar{x},\bar{y}$ space associated with a particular $j$, we can calculate a $j_d$ which agrees with the calculated area.



We show the calculated $j_d$ values as the colour of the simulated trajectories in figure \ref{fig:divingregions}. The agreement between the implied chaotic region and its actual extent is best when there is minimal regular substructure in or below the chaotic region. The actual chaotic region commonly extends to higher $j_{\theta}$ to closer to the poles. At low radii the presence of lots of regular substructure at $j_\theta \rightarrow 0$ leads to a smaller value of $j_d$ than the peak $j$ spanned by the chaotic region.

We will treat $j_d$ as a robust estimator for the rest of this work, but note that a fuller description of the boundary of the chaotic region, and the flux across it, is an interesting avenue of further study. We might expect that the larger the surface area of the chaotic region, the higher the flux into it, and thus our calculation should represent an underestimate.

\subsubsection{$j_d$ across different potentials}

In figure \ref{fig:etad}, we show how $j_{d}$ varies over a range of potentials, by varying $\mu$ and $\alpha$ in our test potential. Each panel corresponds to a single potential, and shows the behaviour over a range of integrals of motion. 

The highest $j_d$ values occur for $R_c$ close to the influence radius. Within the influence radius chaotic behaviour is rapidly suppressed, disappearing for $R_c \lesssim 0.5 r_0$. There is a clear upper envelope at high $j_z$ beyond which chaotic orbits are not seen. For very flattened potentials (high $\alpha$) there can be orbits with $j_z>0.5$ which are still chaotic near the influence radius. At larger energies this critical $j_z$ decreases approximately linearly.


A simple intuition to understand this behaviour is to ask which trajectories manage to pass through the influence radius. This requires an apoapse larger than the influence radius and a periapse within the influence radius. As apoapse occurs at $\lesssim2 R_c$ we see chaos for $\rho \gtrsim 0.5$. Periapse distances generally depend on both $R_c$ and $l$, with smaller $l$ needed to bring a wider orbit to the influence radius. Thus it appears that the driver of chaotic behaviour is trajectories which experience multiple geometries or the transition between them.

In our most spherical potential (high mass ratio and low flattening) we still see a wide range of chaotic behaviour, though even the largest values of $j_d-j_z$ are low ($\sim 0.03$). The requirement that the chaotic region spans at least two cells puts a lower limit on the detectable values ($\gtrsim 0.01$), and it is not clear if chaotic behaviour is eliminated or just highly suppressed as $\alpha \rightarrow 0$.

We also note that there are occasional roughly vertical stripes of regular trajectories in the region we would otherwise expect to be chaotic. Upon inspection these are very radial teardrop-shaped trajectories (in $R$ and $z$) that stay always very close to the equatorial plane, which are only stable at a small subset of energies. It is interesting to note that trajectories in an isolated MN disk potential are mostly not chaotic (with a few exceptions in very particular regions of parameter space; \citealt{Pascale22}), and that our intuition is that it is the fact that trajectories traverse varied geometries that lead to them dispersing chaotically. Thus these regular teardrop orbits may be those that don't experience sufficiently diverse geometries --staying confined to the disk plane which can be locally approximated as spherical-- to disperse.

Overall we see that chaotic orbits are ubiquitous at low $j_z$, values of $j_d$ approaching 1 are seen for flattened potentials near the influence radius, and even much larger orbits or more flattened potentials commonly have $j_d-j_z \gtrsim 0.1$. 

We have also experimented with non-singular spherical potentials (i.e. replacing the MBH by a finite spherical mass distribution) such as one might use for a simple model of the Milky Way bulge and disk, and the results align very well with those from our test potential, giving a similar triangular region of chaotic behaviour, $j_d$ approaching 1 near the radius of transition between disk and bulge dominated regions. Thus we would expect this same level of ubiquity of chaotic low $j_z$ trajectories in any non-spherical potential, regardless of whether they contain a central singularity.


\section{Equilibrium models combining scattering, diving and loss}
\label{sec:eq_models}

We now build a self-consistent equilibrium framework for the probability distribution $p(j_z, j)$ at a fixed orbital energy $\varepsilon$, accounting simultaneously for three physical processes: 

\begin{itemize}
\item (i) two-body scattering, which drives random diffusion in $j$ and $j_z$; 
\item (ii) chaotic angular-momentum redistribution, which for orbits with $j < j_d$ rapidly randomises $j$ between periapse passages independently of scattering; and 
\item (iii) disruption, which removes systems with $j < j_t$ at periapse.
\end{itemize}

We do not model redistribution in $\varepsilon$: energy diffusion is assumed slow compared with angular-momentum diffusion, so we seek an equilibrium in $j_z$–$j$ space at fixed $\varepsilon$. We also treat $j_d$ as independent of $j_z$ for a given energy, a simplification that enables the analytic treatment below.

Although all three processes act simultaneously, we model them as sequential steps applied once per radial orbit. The natural timescale is therefore the orbital period, $\sim T_c(\varepsilon)$, and we apply the steps in the order scattering $\to$ chaotic redistribution $\to$ loss $\to$ renormalisation. Working at periapse is the natural choice: disruptions are defined there, so the state just after the loss step corresponds to the orbit immediately following periapse, with all systems that had $j < j_t$ already removed. The scattering we include is accordingly the orbit-averaged diffusion accumulated over one full radial period. Provided each step is a small perturbation to the total distribution, the outcome is to first order independent of the ordering.

For the chaotic redistribution step we adopt the extreme assumption that diving orbits fully randomise $j$ between consecutive periapses, i.e., that the chaos is strong enough that angular momentum is uncorrelated from one periapse to the next. If a chaotic trajectory instead required $M$ orbits to decorrelate, the contribution of diving orbits to the disruption rate would be reduced by a factor $M$. We use the fully-mixed limit throughout.

We consider the equilibrium probability distribution of systems, $p(j_z, j)$, normalised such that
\begin{equation}
\label{eq:p_norm}
\int_0^1\int_0^j p(j_z,j)dj_z dj=1,
\end{equation}
with $p = 0$ outside the physical region $0 < j < 1$, $j_z < j$. We define $\Gamma$ as the total probability that a system is lost at each step.

\subsection{ Total rate and separability}
\label{sec:numberandrate}

We work at fixed orbital energy $\varepsilon$, writing the probability distribution as $p(j_z, j|\varepsilon)$ and the disruption probability per orbit as $\Gamma(\varepsilon)$. The (infinitesimal) number of systems with energy $\varepsilon$,$z$-angular momentum $j_z$, and total angular momentum $j$ at time $t$ is \footnote{We find this a more natural system to work in than the more commonly used \textit{distribution function}, which satisfies $dN = f(t,\mathbf{x},\mathbf{v})\,d^3\mathbf{x}\,d^3\mathbf{v}$. Expressing $f$ as a function of $(\varepsilon, j_z, j)$ and changing variables requires an extra Jacobian factor — the density of states $g(\varepsilon,j_z,j) \equiv |d^3\mathbf{x}\,d^3 \mathbf{v}/d\varepsilon\,dj_z\,dj|$ — so that $dN = f(\varepsilon,j_z,j)\cdot g(\varepsilon,j_z,j)\cdot d\varepsilon \,dj_z,dj$. Working with $p$ and $n$ as defined in equation 23 is equivalent to absorbing $g$ into the normalisation, which simplifies the bookkeeping considerably.}

\begin{equation}
\label{eq:dN_all}
dN(t,\varepsilon,j_z,j) = n(t,\varepsilon) p(j_z,j|\varepsilon) d\varepsilon dj_z dj,
\end{equation}

where $n(t,\varepsilon)$ is the energy distribution of the population. Disruptions deplete $n$ on the orbital timescale:

\begin{equation}
\label{eq:dn_eps}
\dot{n}(t,\epsilon) \approx -\frac{\Gamma(\varepsilon) n(t,\varepsilon)}{T_c({\varepsilon)}},
\end{equation}

where $T_c(\varepsilon)$ is the period of a circular orbit at energy $\varepsilon$, used as a proxy for the mean orbital period. 

More terms could be included in equation \ref{eq:dn_eps} to model the diffusion in energy space or accretion of new systems. As long as these effects are slow compared to the relaxation to equilibrium in $j_z$, $j$ space the overall solution is unchanged.

The total number and rate of disruption of systems is thus
\begin{equation}
N(t) = \int n(t,\varepsilon) d\epsilon \mathrm{\quad and \quad} \dot{N}(t) = \int \dot{n}(t,\varepsilon) d\epsilon.
\end{equation}

A key consequence of this formulation is that $p(j_z,j|\varepsilon)$ is independent of $n(t,\varepsilon)$: the shape of the distribution in angular momentum evolves on a relaxation timescale much shorter than the depletion of the energy distribution, so we can seek a stationary $p$ at each $\varepsilon$ separately. For conciseness, we drop the conditioning on $\varepsilon$ henceforth.

We seek a fixed-point solution: a distribution $p(j_z,j)$ that is unchanged after one full cycle of scattering, diving, loss, and renormalisation. The diving step fully randomises $j$ within the region $j < j_d$ for each $j_z$, so the equilibrium $p \equiv p_d(j_z)$ must be uniform in $j$ there. Additionally, $p = 0$ in the loss cone at the start of each cycle because the loss step has just emptied it. Our ansatz for the equilibrium form of $p$ is therefore,

\begin{equation}
p(j_z,j)=\begin{cases}
p(j_z,j), \ & j_d<j<1, \\
p_d(j_z), \ & j_t<j<j_d ,\\
0, \ & 0<j<j_t.
\end{cases}
\label{eq:ansatz}
\end{equation}

\paragraph*{Scattering step: $p\to p^S$} After one step of two-body scattering the distribution becomes

\begin{equation}
p^S=\begin{cases}
p(j_z,j)+\delta p^S(j_z,j), \ & j_d<j<1 \\
p_d(j_z)+\delta p^S(j_z,j), \ & j_t<j<j_d \\
\delta p^S(j_z,j), \ & 0<j<j_t,
\end{cases}
\end{equation}
where $\delta p^S(j_z,j)$ is the net probability flux (per unit $j$) scattered into the point $(j_z,j)$ during one step. Its form is determined by the scattering kernel (see, sec. \ref{sec:kernel}) and can be positive or negative.

\paragraph*{Diving step: $p^S\to p^D$.} For any orbit with $j < j_d$ the diving process randomises $j$ uniformly over $[j_z,j_d]$ at each periapse passage. We first define the net flux of probability scattered \emph{into} the diving region of the $j_z$-slice,
\begin{equation}
D p^S(j_z)\equiv \int_{j_z}^{j_d} \delta p^S(j_z,j) dj,
\label{eq:Dp^S}
\end{equation}
which is the net scattering flux that crosses the chaotic boundary $j = j_d$ inward. The total probability in the diving slice before redistribution is then   $\bigl[j_d-\max(j_z,j_t)\bigr]\cdot p_d(j_z)+Dp^S(j_z)$ :
the first term counts the pre-existing equilibrium $p_d(j_z)$, which occupies only $[\max(j_z,j_t)\,j_d]$ (the undisrupted portion of the slice), and the second counts the newly scattered-in probability. The lower limit $\max(j_z,j_t)$ rather than $j_t$ appears because for $j_z > j_t$ the geometric constraint $j \geq j_z$ sets the lower bound of the occupied diving region, not the loss threshold. After chaotic redistribution this total is spread uniformly over the full diving slice $[j_z,j_d]$ (with width $j\_d-j_z$), giving $p^S\to p^D$ where
\begin{equation}
p^D=\begin{cases}
p(j_z,j)+\delta p^S(j_z,j) \ & j_d<j<1 ,\\
\frac{1}{j_d-j_z}\left([j_d-\max(j_z,j_t)]\cdot p_d(j_z)+Dp^S(j_z)\right) \ & 0<j<j_d.
\end{cases}
\end{equation}
\paragraph*{Loss step: $p^D\to p^L$} All systems with $j < j_t$ are disrupted and removed:

\begin{equation}
p^L=\begin{cases}
p(j_z,j)+\delta p^S(j_z,j), \ & j_d<j<1 \\
\frac{1}{j_d-j_z}\left([j_d-\max(j_z,j_t)]\cdot p_d(j_z)+D p^S(j_z)\right), \ & j_t<j<j_d \\
0, \ & 0<j<j_t.
\end{cases}
\end{equation}
\paragraph*{Renormalisation.} Letting $\Gamma$ be the total probability of disruption per step (determined self-consistently below), we restore the normalisation of Equation \ref{eq:p_norm} via 

\begin{equation} 
p = \frac{1}{1-\Gamma}p^L. 
\label{eq:renorm}
\end{equation}
\paragraph*{Self-consistency and the scattering flux.} The fixed-point condition requires that $(1-\Gamma)^{-1}p^L$ reproduces the ansatz of Equation~\ref{eq:ansatz}. Equating the two expressions in the non-diving region ($j > j_d$) gives immediately 
\begin{equation} 
\delta p^S(j_z,j) = -\Gamma \, p(j_z,j), \qquad j_d < j < 1, 
\label{eq:dps} 
\end{equation} 
i.e. in equilibrium the net scattering loss from any non-diving point exactly offsets the fraction $\Gamma$ recovered by renormalisation. Note that the same result is obtained if the four steps are instead applied in the order scattering $\to$ loss $\to$ renormalisation $\to$ diving, confirming that the fixed-point equations are independent of step ordering to first order.  Substituting Equation \ref{eq:dps}  into equation \ref{eq:Dp^S}, and using the equilibrium form $p = p_d(j_z)$ throughout the diving region, yields
\begin{equation}
\label{eq:Dps}
Dp^S(j_z) =\begin{cases}  \left[(j_t - j_z)  - \Gamma \cdot (j_d-j_z)\right] \cdot  p_d(j_z), & \ 0<j_z<j_t \\
-\Gamma \cdot (j_d-j_z)\cdot p_d(j_z), & j_t<j_z<j_d.
\end{cases}
\end{equation}
For $j_z > j_t$ the flux is always negative: there is a persistent drift of probability towards the loss wedge. Within the loss wedge ($j_z < j_t$) the flux changes sign at the critical value
\begin{equation} 
j_{z,c} = \frac{j_t - \Gamma j_d}{1-\Gamma}. 
\label{eq:Dp^S_final} 
\end{equation}
For $j_z < j_{z,c}$ the flux becomes positive: at such low $j_z$ the loss of diving orbits directly into the loss cone is so rapid that scattering must supply a net \emph{inward} flux to sustain equilibrium. Because self-consistency requires $j_{z,c} > 0$, this places an upper bound $\Gamma < j_t/j_d$.

\paragraph*{Disruption rate.} The disruption probability $\Gamma$ equals the integral of $p^D$ over the loss-cone region $j < j_t$:
\begin{equation}
\begin{aligned}
\label{eq:Gamma_int}
\Gamma =& \int_0^{j_t} \int_{j_z}^{j_t} p^D dj dj_z \\
=& \int_0^{j_t} \frac{j_t - j_z}{j_d-j_z}\left[(j_d-j_t)p_d(j_z)+D p^S(j_z)\right]dj_z.
\end{aligned}
\end{equation}
Substituting Equation  \ref{eq:Dps} and solving for $\Gamma$ gives the central result:
\begin{equation}
\label{eq:Gamma_simp}
\Gamma = \frac{\Pi_d}{1+\Pi_d},
\end{equation}
where 
\begin{equation}
\Pi_d = \int_0^{j_t} (j_t-j_z) p_d(j_z) dj_z.
\end{equation}
The quantity $\Pi_d$ is the probability-weighted exposure of the diving population to the loss cone. For $\Pi_d \ll 1$, $\Gamma \approx \Pi_d$, so the rate is set directly by this overlap integral.

The equilibrium is thus fully determined by finding $p(j_z,j)$ and $p_d(j_z)$ that simultaneously satisfy equations \ref{eq:p_norm},  \ref{eq:dps}, \ref{eq:Dps}, and \ref{eq:Gamma_simp} for a given $j_t$, $j_d$, and scattering kernel. 

\paragraph*{Classical limit.} Setting $j_d = j_t$ removes the diving region entirely and recovers the classical loss-cone problem. In this limit $p_d$ and equation \ref{eq:Dps} are ill-defined (the diving slice has zero width), and the disruption rate reduces to
\begin{equation}
\Gamma_{cl} = \int_0^{j_t} Dp_{cl}^S(j_z) dj_z.
\end{equation}
 The renormalisation step introduces a source term proportional to the local probability density, and equations \ref{eq:dps} and \ref{eq:Dps} are therefore analogous to Fokker--Planck equations \citep{Fokker14, Planck17}  with this extra positive-flux term making equilibrium solutions possible. Previous work  \citep[e.g.,][]{Cohn78,Stone18} instead followed the number density directly, yielding a slowly evolving quasi-equilibrium bounded by the outermost $j$ to which the effect of loss had yet to propagate. The fixed-point formulation adopted here sidesteps the sensitivity to initial conditions that such an evolving solution entails. It is not unreasonable to expect relatively long-lived systems to have evolved close to their equilibrium state, making this fixed-point approach a natural starting point.

\subsection{Scattering kernel}
\label{sec:kernel}


The solution presented so far is generic: $\delta p^S$ and $Dp^S$ could be generated for any particular scattering kernel. Now let's define $\Delta(\delta_z, \delta|j_z,j)$, the probability of a system at $j_z, j$ being scattered to $j_z+\delta_z, j+\delta$ (normalized such that $\int_{-\infty}^{\infty}\int_{-\infty}^{\infty} \Delta d\delta_z d \delta =1$).

From this we can calculate
\begin{equation}
\begin{aligned}
\delta p^S = \int_{-\infty}^{\infty}\int_{-\infty}^{\infty} & p(j_z-\delta_z,j-\delta)\Delta(\delta_z,\delta|j_z-\delta_z,j-\delta)\\
&- p(j_z,j)\Delta(\delta_z,\delta|j_z,j) d\delta_z d\delta.
\end{aligned}
\end{equation}

The boundaries of the physical region can either be accounted for by adjusting the kernel (for example by letting $\Delta=0$ for $j+\delta>1$) or by rejecting scatterings that take us out of the physical region such that $\int \int_{\rm unphys.} \Delta d\delta_z d \delta$ is equal to the probability of no scattering event occurring. We will adopt the latter hereafter, as it is necessary to make the greatly simplifying assumption that $\Delta$ is independent of $j_z,j$. In this case we get the much simpler
\begin{equation}
\begin{aligned}
\delta p^S = \int \int_{\rm phys.} \left[p(j_z-\delta_z,j-\delta)- p(j_z,j)\right]\Delta(\delta_z,\delta) d\delta_z d\delta.
\end{aligned}
\end{equation}

In most physical situations we expect $\Delta$ to have a finite effective width (such that only perturbative changes of state are likely) in which case we \textit{may} be able to locally expand $p(j_z-\delta_z,j)$ around $j_z,j$ giving
\begin{equation}
\begin{aligned}
\label{eq:dps_expand}
\delta p^S \approx& -\langle \delta_z\rangle\partial_{j_z}p-\langle \delta\rangle\partial_{j}p\\
&+\frac{\langle \delta_z^2\rangle}{2}\partial_{j_z}^2p+\frac{\langle \delta^2\rangle}{2}\partial_{j}^2p+\langle \delta_z \delta\rangle\partial_{j_z}\partial_{j}p +\mathcal{O}^3
\end{aligned}
\end{equation}
where all function of $p$ are evaluated at $j_z,j$ and
\begin{equation}
\langle\delta_z^a \delta^b\rangle = \int \int_{\rm phys.} \delta_z^a \delta^b \Delta(\delta_z,\delta) d\delta_z d\delta,
\end{equation}
which, far from any boundaries, correspond to the mean, variance and covariance of $\Delta$. \textit{However}, there are some boundaries (most notably $j=j_d$) where $p$ does not vary smoothly and continuously, and $\delta p^S$ must be calculated more directly.



If $\Delta$ is symmetric (zero mean in both $\delta_z$ and $\delta$) and separable, then $\langle\delta_z\rangle = \langle\delta\rangle = \langle\delta_z\delta\rangle = 0$. In the simplest isotropic case $\langle \delta_z^2\rangle =\langle\delta^2\rangle \equiv \sigma^2$, which sets the single scale of the scattering kernel. For relatively small $\sigma$ the equilibrium $p(j_z,j,|,j_t,j_d,\sigma)$ is moreover largely independent of the functional form of $\Delta$: since the effective kernel over many steps is a convolution of $\Delta$ with itself many times, which tends towards a Gaussian regardless of the shape of $\Delta$ (by the central limit theorem). Equation\ref{eq:dps_expand} then reduces, far from any boundary, to 

\begin{equation}
\delta p^S \approx \frac{\sigma^2}{2}\nabla^2 p +\mathcal{O}^3.
\label{eq:fp}
\end{equation}
The Fokker–Planck approximation, equation\,\ref{eq:fp}, is valid only in the bulk, away from boundaries where $p$ may be discontinuous. To compute the flux at the chaotic boundary $j = j_d$ — which determines $Dp^S$ — we must calculate it more directly.



\subsubsection{Net flux through a boundary}
We can calculate the flux over a 1D boundary, assuming that we're far from the edges of the physical region. We work in coordinates $w$, perpendicular to the boundary at $w=0$, and allow discontinuous solutions $p_\pm(w)$ either side of it. Thus the flux from positive to negative $w$ is
\begin{equation}
D_- = \int_0^\infty \int_{-\infty}^{-w} p_+(w) \Delta(\omega) d \omega d w = \int_0^\infty p_+(w) F_{\Delta}(-w) d w
\end{equation}
where 
\begin{equation}
F_\Delta(w) = \int_{-\infty}^w \Delta(\omega) d \omega
\end{equation}
is the cumulative distribution function (CDF) of $\Delta$. Expanding $p_+(w)=p_{+0}+w\partial_wp_{+0}+\mathcal{O}(w^2)$ where subscript $+0$ denotes the function evaluated approaching $0$ from above. Substituting this in and integrating by parts we get
\begin{equation}
D_- = \sigma K p_{+0} + \frac{\sigma^2}{4}\partial_w p_{+0} +\mathcal{O}(\sigma^3)
\end{equation}
where 
\begin{equation}
K = \frac{1}{\sigma} \int_{-\infty}^0 F_\Delta(w) dw,
\end{equation}
is the area of the CDF for $w<0$. $K$ takes remarkably consistent values regardless of the choice of $\Delta$\footnote{For a Gaussian $K = \frac{1}{\sqrt{2\pi}} \approx 0.399$, whilst for a polynomial kernel\footnote{Any kernel of the form $\Delta(w|n) = A_n\left(1-\left(\frac{|x|}{\chi_n}\right)^n \right)$ normalised such that $A_n=\frac{n+1}{2n\chi_n}$ and $\chi_n=\sqrt{3\frac{n+3}{n+1}}\sigma$} of order $n$, $K_n=\frac{\sqrt{3(n+1)(n+3)}}{4(n+2)}$, which ranges from a maximum of $K_\infty = \frac{\sqrt{3}}{4}\approx 0.433$ for a flat top-hat function, to $K_1=\frac{1}{\sqrt{6}} \approx 0.408$ for a shark-tooth function, to a minimum of $K_0 = \frac{3}{8}=0.375$ for a sharp centrally peaked function.}.
Performing the same analysis for flux from negative to positive gives similarly,
\begin{equation}
D_+ = \sigma K p_{-0} - \frac{\sigma^2}{4}\partial_w p_{-0} +\mathcal{O}(\sigma^3).
\end{equation}
Thus the net flux over any such boundary, from positive to negative, is
\begin{equation}
\begin{aligned}
\label{eq:D_net_general}
D =& D_-- D_+ \\
=& \sigma K (p_{+0}-p_{-0}) + \frac{\sigma^2}{4}(\partial_w p_{+0}+\partial_w p_{-0}) +\mathcal{O}(\sigma^3).
\end{aligned}
\end{equation}
If the boundary has unit normal $\mathbf{\hat{w}}$ then $\partial_{|\mathbf{w}|} p = \mathbf{\hat{w}} \cdot \bm{\nabla}p$. Thus we see that when there is a discontinuity there is a usually dominant first order (in $\sigma$) flow over that boundary, but without a discontinuity there is still net flow with direction and magnitude equal to $-\frac{\sigma^2}{2} \bm{\nabla} p$.

The boundary of most interest to us is the regular/chaotic boundary at $j=j_d$ (with $\mathbf{\hat{w}}=\mathbf{\hat{j}}$ and $|\mathbf{w}|=j-j_d$) where equation \ref{eq:D_net_general} gives a net flux \textit{into the diving region} of
\begin{equation}
\label{eq:D_net}
D = \sigma K (p(j_z,j_d)- p_d(j_z)) +\frac{\sigma^2}{4}\partial_jp(j_d,j_z) 
\end{equation}
where we expect both terms to be positive as there should be net scattered flux into the diving region and towards the loss cone. Note that equation~\ref{eq:D_net} accounts only for scattering flux across the chaotic boundary; it does not include drift between $j_z$-slices within the diving region, and so represents just one component of $Dp^S$.

\subsection{Approximate 2D solution}
\label{sec:approx_2d}
We now seek an explicit approximate form for $p$ in the non-diving region ($j_d < j < 1$), where two-body scattering dominates and the self-consistency condition equation~\ref{eq:dps} applies. Substituting the Fokker--Planck approximation for the scattering flux, equation~\ref{eq:fp}, into equation~\ref{eq:dps} gives $\tfrac{\sigma^2}{2}\nabla^2 p = -\Gamma p$  which rearranges to 
\begin{equation}
\label{eq:dps_approx}
\nabla^2 p = -\gamma^2 p,
\end{equation}

with 
\begin{equation} 
\gamma^2 = \frac{2\Gamma}{\sigma^2}, 
\label{eq:gamma_def}
\end{equation}
a slight variation on the heat equation that occurs for diffusion problems. Here $\gamma \geq 0$ is an inverse length scale in angular momentum space that sets the curvature of $p$  --it grows with the loss fraction $\Gamma$ and shrinks with the diffusion coefficient $\sigma$. 

We note, however, that equation ~\ref{eq:dps_approx}  is valid only away from the chaotic boundary. In the boundary layer $j_d < j \lesssim j_d + \sigma$ the Fokker--Planck approximation does not capture the extra scattering flux into the diving region, so the equation breaks down there and we cannot assume $p \approx p'$. This makes it impossible to compute $Dp^S$ self-consistently near $j = j_d$ and hence to determine the amplitude and curvature of the solution analytically; we return to this point below.

The boundary conditions (BCs) of equation \ref{eq:dps_approx}  are set by the fact that the diffusion flux is proportional to $\bm{\nabla} p $ , so wherever we expect no net probability flux across a boundary of the physical region the solution must have zero gradient perpendicular to that boundary.

\begin{equation} \partial_{j_z} p\big|_{j_z=0} = \partial_j p\big|_{j=1} = \bigl(\partial_j p - \partial_{j_z} p\bigr)\big|_{j=j_z}= 0. 
\label{eq:BCs} \end{equation}
The first condition follows from the symmetry of the problem in $l_z$: since $j_z \propto |l_z|$, the distribution must be even in $l_z$, so that there is no net flux through the $j_z = 0$. (If a scattering would give $j_z + \delta_z < 0$ the system instead lands at $|j_z + \delta_z|$, a reflection best accounted for by imposing this BC). The second condition imposes no flux through the outer boundary $j = 1$. The third condition enforces no flux across the diagonal boundary $j = j_z$, the physical lower edge of the domain where $j$ reaches its minimum value for a given $j_z$; the combination $\partial_j p - \partial_{j_z} p$ is the component of $\nabla p$ normal to this boundary. There is a fourth, more complex boundary condition, essentially expressed by equation \ref{eq:Dps}, that the diffusion near, into, and within the chaotic boundary be self-consistent with the variation in probability density therein.

As the third and fourth BCs cannot be expressed as independent expressions in $j_z$ and $j$, there is no separable solution in these coordinates. However, a useful approximate solution can be found by working in terms of a radius in $(j_z,j)$ space,
\begin{equation}
r_j^2 =j_z^2 + j^2,
\end{equation}
and corresponding angle $\theta_j$.
The first and third BCs correspond to constant $\theta_j$, and there is no gradient of $r_j$ (or functions derived thereof) normal to these boundaries. Along the second boundary ($j = 1$) we have $r_j = \sqrt{j_z^2 + 1}$, which varies from 1 to $\sqrt{2}$ as $j_z$ goes from 0 to 1. Whilst the second BC does not correspond to a true constant value of $r_j$, the variation is relatively small, and we are most interested in the form of the solution for $j_z \ll 1$. 

Thus we make the ansatz $p \approx p'(r_j)$, which satisfies the first and third BCs exactly, and the second approximately. In 2D the Laplacian reduces to
\begin{equation}
\frac{1}{r_j} \partial_{r_j} (r\partial_{r_j}p')=-\gamma^2 p',
\end{equation}
which takes solutions of the form of Bessel functions of the first kind:
\begin{equation}
p'(r_j) = A J_0(\gamma r_j) + B Y_0 (\gamma r_j).
\end{equation}
It is interesting to note that omitting the renormalisation term (letting $\gamma = 0$) gives solutions of the form $p'(r_j) \propto \ln(r_j)$ --- the typical form seen in, for example, \citet{Merritt13book}.
The gradient, with respect to $r_j$ is
\begin{equation}
\partial_{r_j} p' = - \gamma \left(A J_1(\gamma r_j) + B Y_1(\gamma r_j) \right).
\end{equation}
Since we expect $p \geq 0$ to be monotonically increasing towards the outer boundary, we require $\partial_{r_j} p' \geq 0$ throughout the domain. Applying the no-flux condition at the outer boundary, approximated at $r_j = \sqrt{2}$ (the corner $j_z = j = 1$), gives 
\begin{equation}
B= -A \frac{J_1(\sqrt{2}\gamma)}{Y_1(\sqrt{2}\gamma)}.
\end{equation}
For $B \geq 0$ we need $Y_1(\sqrt{2} \gamma) \leq 0$. The function $Y_1(x)$ is negative for $x$ less than its first positive zero $x_1 \approx 2.20$, so this restricts $\gamma \leq x_1/\sqrt{2} \approx 1.56$. A further upper limit $\gamma_{ \rm max}(j_d)$ comes from requiring the distribution to be non-negative at the chaotic boundary, $p'(0,j_d) \geq 0$, which is satisfied by
\begin{equation}
\label{eq:gamma_max}
\frac{J_0(\gamma_{\rm max} j_d)}{J_1(\sqrt{2}\gamma_{\rm max})} = \frac{Y_0(\gamma_{\rm max} j_d)}{Y_1(\sqrt{2}\gamma_{\rm max})}
\end{equation}
The physically admissible range is therefore $0 < \gamma \leq \gamma_{\rm max}(j_d)\lesssim 1.56$.
The values of  $\gamma$ and $A$ within this range must still satisfy the normalisation condition and self-consistency with the net flow rate across the chaotic boundary. Because the boundary-layer region prevents a fully analytic determination, we adopt the following pragmatic conditions instead.

In the classical case ($j_d \to j_t$) the diving region vanishes, so we require $p'(0,j_t) = 0$ and accordingly take $\gamma = \gamma_{\rm max}$. 

In the diving case the probability density at the chaotic boundary is not negligible. Lacking an analytic form for $p_d(j_z)$ near this boundary, we use the numerical equilibrium solutions of section \ref{sec:toy_model} to motivate a functional form: $p_d(j_z)$ is found to be approximately quadratic for $0 < j_z < j_t$ and linear for $j_t < j_z < j_d$. At the far corner of the diving region, $j_z = j = j_d$, the diving population is negligible, so chaotic redistribution cannot advect systems away from there, and hence we assume $p_d(j_d) \approx p'(j_d,j_d)$ and $\partial_{j_z} p_d(j_d) \approx \partial_{j_z} p'(j_d,j_d)$. Matching these conditions with a self-consistent loss rate and normalisation specifies a unique solution. 

\begin{figure}
\includegraphics[width=\columnwidth]{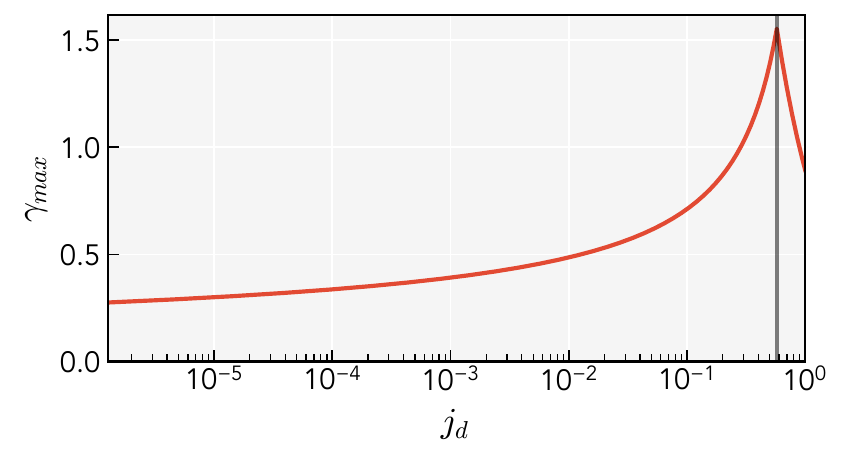}
    \caption{The maximum curvature parameter, $\gamma$, as given by equation \ref{eq:gamma_max}, such that the approximate solution $p'\geq 0$ for all $j\geq j_d$. It reaches a peak (of $\sim 2.20/\sqrt{2}$) at $j_d \sim 0.575$ (vertical line), corresponding to the highest possible curvature solution with non-negative gradients.}
    \label{fig:gamma_max}
\end{figure}

The existence of $\gamma_{\rm max}(j_d)$ implies a corresponding $\Gamma_{\rm max}(j_d) = \tfrac{1}{2}\sigma^2\gamma_{\max}^2$ (from equation~\ref{eq:gamma_def}), which provides an approximate upper limit on the disruption rate of any system that has fully relaxed to an equilibrium state. We show $\gamma_{\rm max}(j_d)$ in Figure \ref{fig:gamma_max}. It is not clear from our numerical experiments whether it asymptotes to a particular value or scaling relationship as $j_d \to 0$, but it remains $\gtrsim 0.1$ even for very small $j_d$. There is a peak near $j_d \approx 0.575$, beyond which the constraint $\partial_{r_j} p \geq 0$ limits the curvature. The fact that $\gamma_{\rm max}^2 \gtrsim 0.1$ suggests that $\Gamma_{\rm max}$ should be within an order of magnitude of $\sigma^2$ for essentially all $j_d$.

Although this approximate solution does not capture the boundary layers associated with the full set of boundary conditions, it provides a useful analytical description of the equilibrium distribution and a benchmark for the numerical results presented in the next section. In particular, it captures the large-scale structure of the distribution sufficiently well to provide an effective initial condition for the simulations, substantially reducing the time required to reach equilibrium. We therefore use this approximate solution throughout Section \ref{sec:toy_model} as the starting point for our numerical calculations.

\subsection{Analytic 1D solution}
\label{sec:1d}

The expressions for the equilibrium distribution, \ref{eq:p_norm}, \ref{eq:dps}, \ref{eq:Dps} and \ref{eq:Gamma_simp} can be written down for a 1D distribution of $p_z(j)$. Although this  corresponds to the analogous (but unphysical) case of redistribution of $j$ but not $j_z$, keeping $j_z$ fixed allow us to find an analytical solution against which, in Section \ref{sec:1D_model} , we  test our  Monte Carlo numerical calculation detailed in  Section \ref{sec:toy_model}.

The equivalent probability density now satisfies
\begin{equation}
\int_{j_z}^{1}p_z(j) dj =1,
\label{eq:norm_1D}
\end{equation}
with the lower limit of integration set by the geometrical constraint $j \ge j_{z}$, and is of the form
\begin{equation}
p_z(j) = \begin{cases}
p_z(j), \ & j_d<j<1 ,\\
p_{d,z}, \ & j_t<j<j_d, \\
0, \ & j_z<j<j_t.
\end{cases}
\end{equation}
Note that in cases with $j_z>j_t$ there is no loss and the equilibrium solution is trivial, $p_z=\frac{1}{1-j_z}$.

Far from any boundary, the 1D Fokker–Planck for scattering in $j$ gives $\delta p^S \approx \frac{\sigma^2}{2}\partial_j^2 p_z$. At equilibrium this must balance the loss: $\partial_j^2 p_z = -\gamma_z^2 p_z$. This is the 1D analogue of eq. \ref{eq:dps_approx}, but with much simpler trigonometric solutions. Enforcing that $\partial_j p_z \geq 0$ always, with no gradient and hence no flux over the $j=1$ boundary gives the general form:
\begin{equation}
\label{eq:p_1d}
p_z=A_z \cos\left(\gamma_z(1-j)\right),
\end{equation}
which, after applying the normalization condition (eq. \ref{eq:norm_1D}) specifies
\begin{equation}
A_z= \frac{\gamma_z\left( 1-p_{d,z}(j_d-j_t)\right)}{\sin\left(\gamma_z(1-j_d) \right)}.
\end{equation}
Since we expect $p_z \geq 0$ and $A_z > 0$ , we should impose $\cos(\gamma_z(1-j)) \geq 0$ throughout $j \in [j_d, 1]$. The argument $\gamma_z(1-j)$ is largest at $j = j_d$, so the constraint is: $\gamma_z(1-j_d) \leq \frac{\pi}{2}$ giving: $\gamma_{z,\rm max} = \frac{\pi}{2(1-j_d)}$.

Finally,  using eq \ref{eq:Gamma_simp},  the loss fraction is:
\begin{equation}
\label{eq:Gamma_1d}
\Gamma_z = \frac{\sigma^2 \gamma_z^2}{2} = \frac{(j_t-j_z)p_{d,z}}{1+(j_t-j_z)p_{d,z}}.
\end{equation}
There is a unique solution for any given scattering kernel that matches the net scattered flux across the boundary, $D_z^S$, to $\Gamma_z$. We can write down 1D versions of equations \ref{eq:Dps} and \ref{eq:D_net_general} giving,
\begin{equation}
\begin{aligned}
D_z^S =& \left[ (j_t-j_z) - \frac{\sigma^2 \gamma_z^2}{2} (j_d-j_z)\right]p_{d,z} \\
=& K\sigma(p_z(j_d)-p_{d,z}) + \frac{\sigma^2}{4}\partial_j p_z(j_d),
\end{aligned}
\end{equation}
which, along with equations \ref{eq:Gamma_1d} and \ref{eq:p_1d}, is only satisfied by a single value of $0 \leq \gamma_z(j_z,j_t,j_d,\sigma) \leq \gamma_{z,\rm max}$.

We show the agreement between this analytic solution and a Monte Carlo model in section \ref{sec:1D_model}.

\section{Toy models}
\label{sec:toy_model}

To examine the behaviour in detail, we can construct walker based Monte-Carlo toy models\footnote{A minimal working example of which can be found at \url{https://github.com/zpenoyre/losswedge}.} which broadly replicate the physics of the combined formalism, but which are also simple to interpret and relatively numerically inexpensive.

The scattering step is performed by incrementing $j_z$ and $j$ by some random $\delta j_z$ and $\delta j$, both of which are drawn from a symmetric distribution with variance $\sigma^2$. 
As the final distribution is the result of many scatterings, we expect the net result to tend to a normal distribution regardless of the choice of scattering kernel, $\Delta$. This is because the net result of many events is the convolution of each $\Delta$, and a Gaussian is stable under self-convolution.

Some care must be taken for scatterings near the physical boundaries of the space ($0\leq j\leq 1$ and $j_z\leq j$). We reject (and do not resample) any step that would take a system out of the valid region. Without loss this leads to a uniform equilibrium distribution.

For some system at ($j_z,j$),  we perform these operations at each step:
\begin{enumerate}
\item SET $\delta j \sim \mathcal{N}(0,\sigma^2)$, $\delta j_z \sim \mathcal{N}(0,\sigma^2)$, \\
IF $0<j+\delta j < 1$ AND $|j_z +\delta j_z|<j+\delta j$ \\
SET $j=j+\delta j$, $j_z=|j_z+\delta j_z|$

\item IF $j<j_d$ \\
SET $j\sim \mathcal{U}(j_z,j_d)$

\item IF $j<j_t$  \\
CHOOSE any other non-disrupted system at ($\hat{j}_z,\hat{j}$) \\
SET $j=\hat{j}$, $j_z=\hat{j}_z$
\end{enumerate}
where $\mathcal{N}$ and $\mathcal{U}$ represent random draws from a Gaussian and uniform distribution respectively.

In the first step the scattering can lead to $j_z+\delta j_z<0$, but as $j_z$ an absolute value we take the magnitude.

The last step here corresponds to a disruption, and then the subsequent redistribution of those systems randomly, weighted by the probability density of surviving systems. $\Gamma$ is then equal to the fraction of systems disrupted at each step.

\subsubsection{1D toy model}
\label{sec:1D_model}

We have an exact numerical solution for a 1D formalism, as derived in section \ref{sec:1d}, where systems diffuse in $j$ but not $j_z$. We can construct an equivalent numerical formulation of this situation and show that the two agree.

\begin{figure}
\includegraphics[width=\columnwidth]{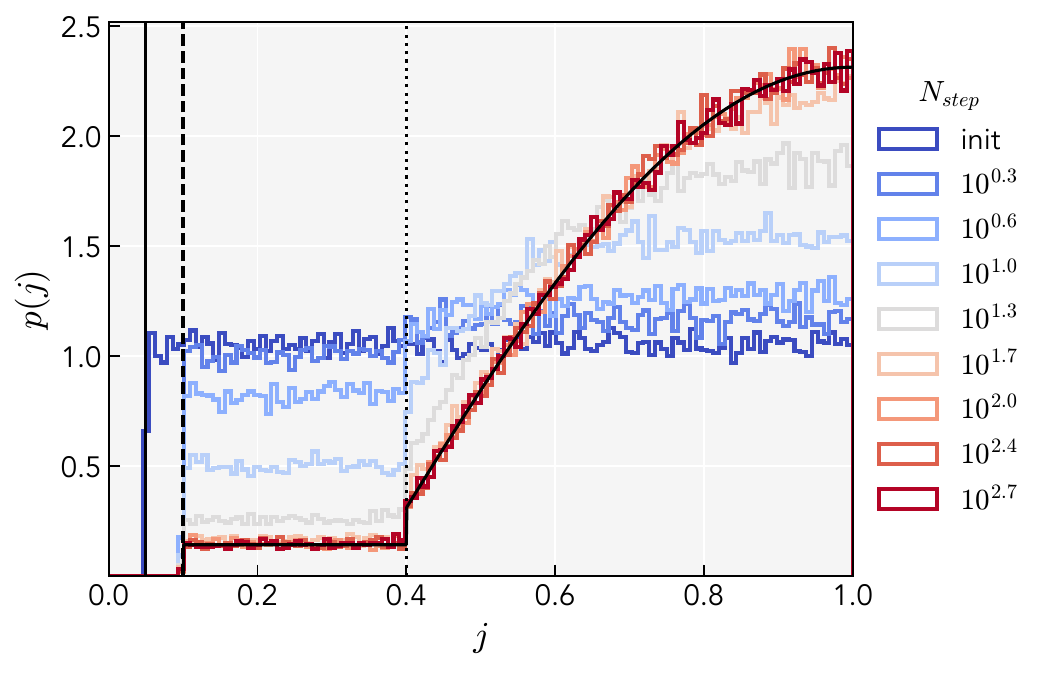}
    \caption{An example of evolving a 1D distribution of $p(j|j_z,j_t,j_d,\sigma)$ using a Gaussian scattering kernel, with $j_d=0.4$ (dotted line), $j_t=0.1$ (dashed line), $j_z=0.05$ (solid line) and $\sigma=0.1$. We start with an initially uniform distribution, which after around a hundred steps reaches a static equilibrium which it maintains thereafter. The analytic equilibrium solution is over-plotted in black.}
    \label{fig:1dexample}
\end{figure}

In Figure \ref{fig:1dexample} we show the evolution of $p(j)$ in 1D, starting from initially uniform walkers and evolving after around 100 steps to the equilibrium distribution, which it then maintains thereafter. The analytic solution is also shown, and agrees excellently with the numerical equilibrium state.

This isn't a truly physical state as it suppresses any change in $j_z$, from scattering or refilling, but it does neatly validate both our numerical and analytic formalism, and allows us to return with confidence to the more complex 2D case.

\subsection{Convergence}

We wish to stop our simulations when they have reached a stable equilibrium state. This should correspond to a steady value of $\Gamma$, and a relatively constant $p(j_z,j)$. In practice this is a tricky condition to reliably test, as the evolution can be quite slow (especially for $\sigma \ll 1$) and numerical noise can cause variations that temporarily mimic a steady state. Thus we cannot rely on an apparently steady value of $\Gamma$ alone.

It is useful to split the physical region into four subregions:
\begin{itemize}
\item N - $j>j_d$
\item D - $j<j_d$ \& $j_z>j_t$
\item W - $j<j_d$ \& $j_z<j_t$
\item C - $j<j_t$
\end{itemize}
where we note that C, the loss cone, is a subset of W.

Walkers in N only experience scattering. In D they also experience diving orbits, but these are not capable of bringing them into C. This differs from the loss wedge, W, where chaotic redistribution can deposit walkers into C.

We can denote the total probability, $P_A$, as the integrated fraction of all walkers in some region $A$.

We can also define the fraction of walkers, $\Gamma_{AB}$, moving from region $A$ to $B$ via scattering at each step. There is also a fraction of walkers moving into each region every step via redistribution of lost walkers, $\Gamma_{CA}$. Often the net flow across a boundary, $\bar{\Gamma}_{AB}=\Gamma_{AB}-\Gamma_{BA}$, is the more pertinent quantity, and can be much smaller than $\Gamma_{AB}$ or $\Gamma_{BA}$.

With these definitions we can write down the net flow into the W region
\begin{equation}
\label{eq:Gamma_W}
\Gamma_W = \bar{\Gamma}_{NW} + \bar{\Gamma}_{DW} + \Gamma_{CW}
\end{equation}
and the net flow into the combined W and D regions
\begin{equation}
\Gamma_{(WD)} = \bar{\Gamma}_{NW} + \bar{\Gamma}_{ND} + \Gamma_{CW} + \Gamma_{CD}.
\end{equation}

We expect the net flow to be from large $j_z$ and $j$ to small, and thus from N, to D and W, and into C. Thus in equilibrium
\begin{equation}
\label{eq:Gamma_check}
\Gamma=\Gamma_W=\Gamma_{(WD)}.
\end{equation}
This is a useful condition because it links the behaviour at small scales, $j\sim j_t$, (encoded into $\Gamma$) to larger scales, $j\sim j_d$ (encoded into $\bar{\Gamma}_{NW}$ and $\bar{\Gamma}_{ND}$).

Especially for low $\Gamma$ any of these quantities can be dominated by random noise step-to-step. Thus we calculate $\Gamma$, and it's associated uncertainty (including autocorrelated noise) $\sigma_\Gamma$, over a window containing the most recent one third of steps. From this we can check if equation \ref{eq:Gamma_check} is satisfied. We also check if $\Gamma$ and $P_N$ are consistent with the values calculated, over their respective window, one-third of steps ago. If all these conditions are satisfied the simulation is considered to have reached equilibrium.

\subsection{Initial conditions}
\label{sec:toy_ic}

\begin{figure}
\includegraphics[height=0.85\textheight]{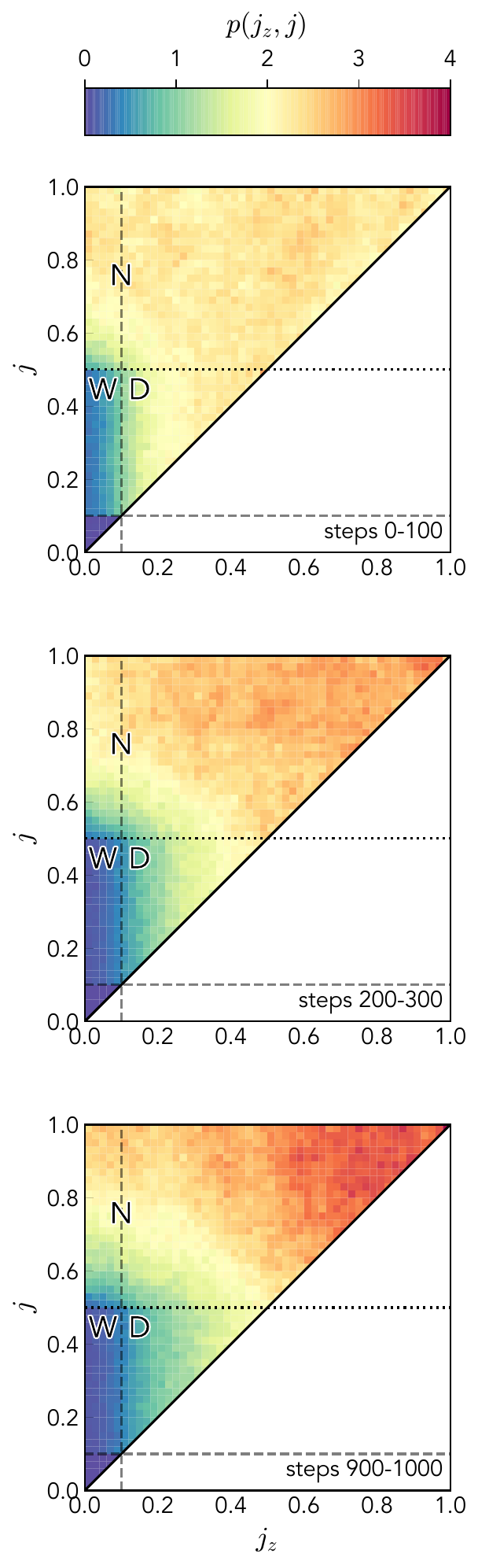}\\
    \caption{Evolution from an initially flat $p$ for 10,000 walkers with $j_t=0.1$ (dashed line) and $j_d=0.5$ (dotted line) and $\sigma=0.02$. The solid black line denotes the edge of the physical region ($j>j_z$). The N, D and W regions are denoted. We average the occupation fractions over 100 steps, though there is still some visible Poisson noise (particularly in the D and W regions where we know there should be no vertical gradient of $p$).}
    \label{fig:p_evolve}
\end{figure}

The simplest possible initial conditions (ICs) are flat, with $p=2/(1-j_t^2)$ everywhere in the physical region. In figure \ref{fig:p_evolve} we show the evolution of $p$, from flat ICs, towards an equilibrium state.

At early times (first panel) we see that the loss wedge (W) is quickly depleted, most strongly for $j_z\rightarrow 0$, where chaotic redistribution is most likely to deposit a system into the loss cone. 

At slightly later times (second panel) the whole diving region shows significant depletion, as well as N close to the loss wedge. Far from the loss cone the distribution is still relatively flat, though the probability is enhanced overall. This is because disrupted systems have been refilling this region, but there has not yet been enough time for a self consistent gradient (and hence flux) to develop here.

At late times (third panel), once equilibrium is reached, there is a consistent gradient throughout the physical region.

As net diffusion both requires and establishes a gradient, any region that shows significant depletion quickly attains a near equilibrium state (W and D do not evolve much after the first and second panel respectively). Conversely the lack of an initial gradient far from the loss cone, and the comparatively large distances walkers there need to traverse, leads to a very slow asymptote to global equilibrium.

\begin{figure}
\includegraphics[width=0.9\columnwidth]{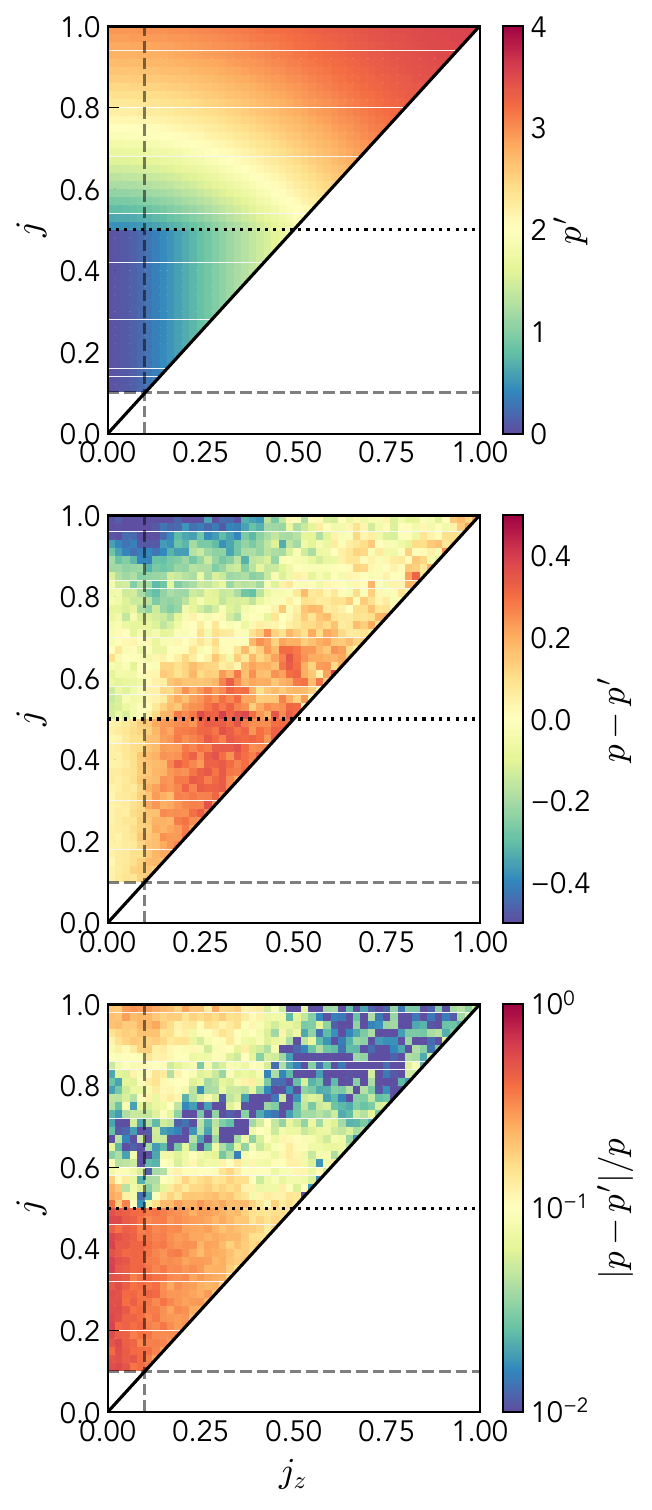}\\
    \caption{A comparison of the equilibrium solution, $p$ (last panel of figure \ref{fig:p_evolve}) with $p'$, the approximate equilibrium solution which we use as initial conditions for all future simulations, as described in section \ref{sec:toy_ic}. The top panel shows $p'$ itself, the second panel shows the absolute difference, and the final panel shows the relative difference.}
    \label{fig:p_init}
\end{figure}

In section \ref{sec:approx_2d} we introduced an approximation for the equilibrium solution, $p'$. In figure \ref{fig:p_init} we compare the numerical equilibrium, which we treat as the true solution, $p$, to $p'$. The two match well, to within $10\%$ of the true value throughout most of N. One of the largest deviations occurs as $j_z\rightarrow 0$ and $j\rightarrow 1$, which is to be expected as we know that our approximate solution doesn't perfectly match the gradient BCs here. There is also significant deviation in the region just above $j=j_d$, where we expect flow across $j_d$ to create a boundary layer not captured by our approximate solution, which also leads to significant deviations for $j<j_d$.

Thus $p'$ represents a moderately good approximation of the equilibrium solution. More than that, it is an excellent initial condition to start the simulations from. It fits very well far from the loss cone, where the evolution is slowest, and where it deviates most (close to and below $j=j_d$) the gradients are large and walkers can quickly redistribute.

\begin{figure}
\includegraphics[width=0.8\columnwidth]{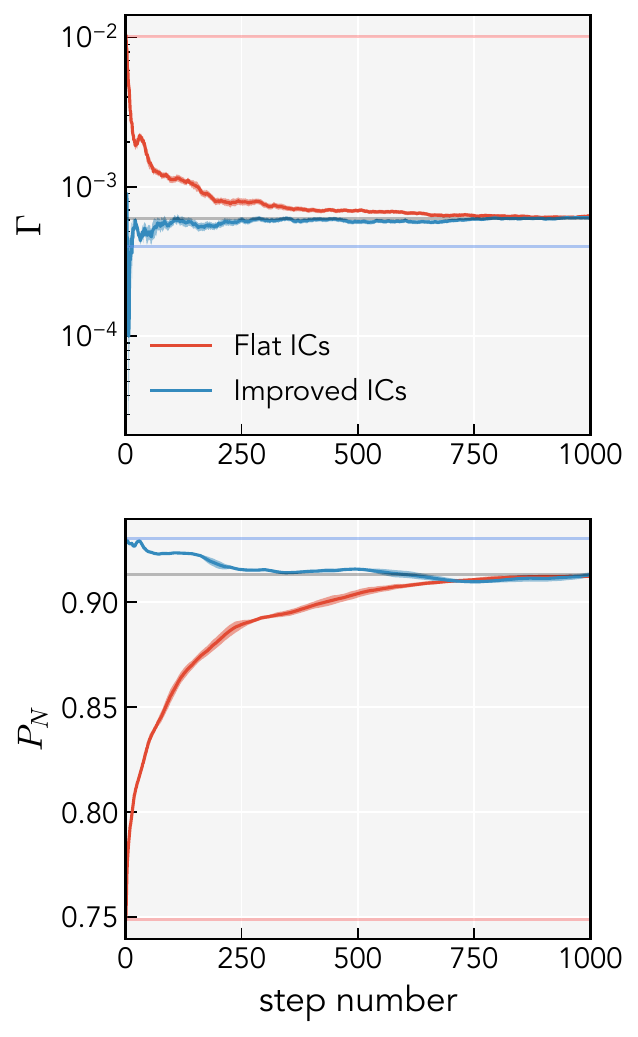}\\
    \caption{The evolution of $\Gamma$ and $P_N$ for the initially flat simulation shown in figure \ref{fig:p_evolve} (red) as well as the improved initial conditions as discussed in section \ref{sec:toy_ic} (blue). The initial values for each are shown as coloured horizontal lines, and the line value in grey.}
    \label{fig:p_Gamma_N}
\end{figure}

We can see the difference in the convergence in figure \ref{fig:p_Gamma_N}, where we show how $\Gamma$ and $P_N$ evolve over time, both for flat ICs and using  $p'$ as ICs. Both reach the same equilibrium, but $p'$ starts significantly closer and quickly converges (after a few hundred steps), whilst the flat ICs asymptote to equilibrium much more slowly (requiring around 1000 steps). Given that rate of convergence depends strongly on $\sigma$, these improved ICs allow us to simulate significantly smaller $\sigma$, and we will use them throughout the rest of the work.

\subsection{Recovering the classical case}
\label{sec:classical}

By setting $j_d=j_t$ we essentially eliminate the effect of diving orbits, as regardless of what occurs in step (ii) all chaotically redistributed systems are disrupted in step (iii). Thus we recover the classical case, and should expect to recover the classical results for $\Gamma_{cl}(j_t,\sigma)$.

Following \citet{Merritt13book} we can split the classical behaviour into two cases, depending on the magnitude of
\begin{equation}
q=\frac{P \mathcal{D}}{j_t^2}
\end{equation}
where $P$ is the period of an orbit (of a given energy) and $\mathcal{D}$ is the orbit averaged diffusion coefficient, recording the rate of diffusion in $j^2$. 

Using \citet{Merritt13book} equations 6.28 and 5.167b we can write
\begin{equation}
\mathcal{D}=\lim_{j\rightarrow0}\frac{\langle (\Delta j^2)^2\rangle}{2j^2}\approx 2 \langle (\Delta j)^2 \rangle
\end{equation}
where the middle equality is approximate as it ignores (comparatively minor) diffusion in energy. 

Rewriting in terms of our model, which calculates deviations in $j$ once per orbit, $\langle (\Delta j)^2\rangle = \sigma^2/P$ and hence
\begin{equation}
q=\frac{2 \sigma^2}{j_t^2}.
\end{equation}

When $q\gg 1$ (the \textit{full} regime) walkers take random steps much larger than the size of the loss cone each step, and the fraction of all systems of a given energy that are disrupted over one period is
\begin{equation}
\label{eq:Gamma_cl_full}
\Gamma_{cl} \sim \frac{\dot{n}(t,\varepsilon)}{n(t,\varepsilon)} T_c(\epsilon) \sim j_t^2
\end{equation}
(\citealt{Merritt13book} equation 6.11).

When $q\ll1$ (the \textit{empty} regime) diffusion into the loss cone is slow, and there is significant depletion of orbits with $j\gtrsim j_t$. In this case fraction of systems disrupted per orbit is
\begin{equation}
\Gamma_{cl} \approx \frac{q j_t^2}{|2\ln j_t|} \approx \frac{\sigma^2}{|\ln j_t|}
\end{equation}
(\citealt{Merritt13book} equation 6.38).

\begin{figure}
\includegraphics[width=0.8\columnwidth]{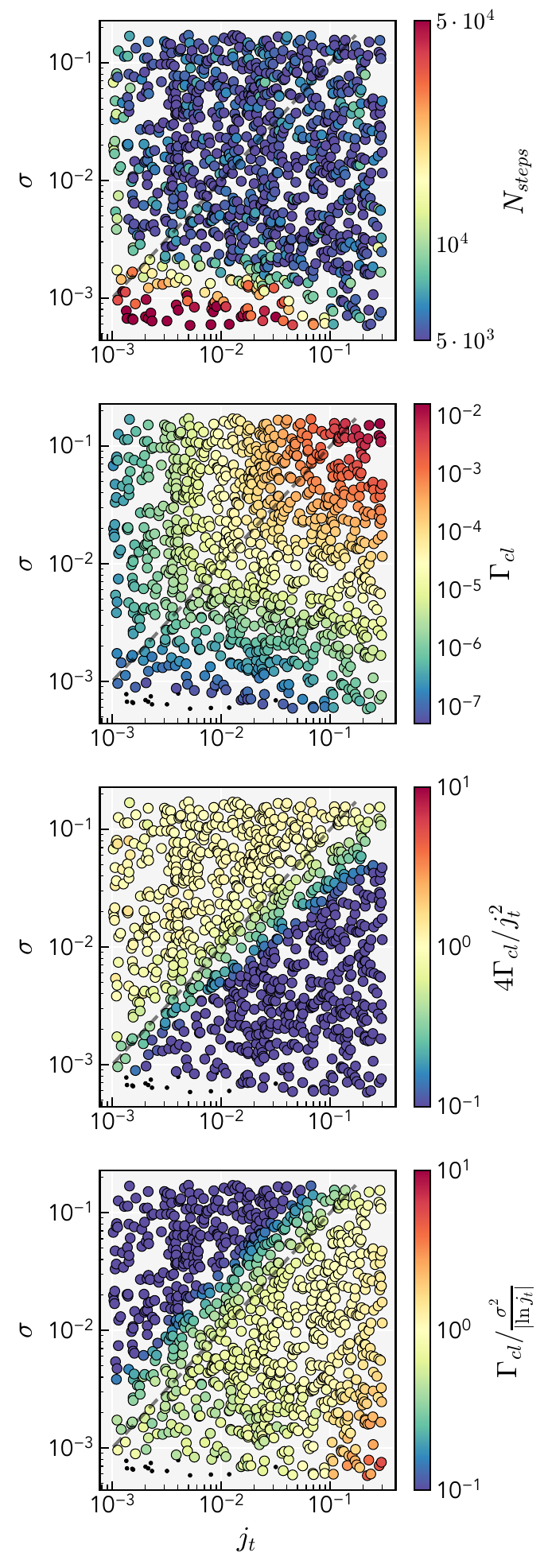}\\
    \caption{Testing the results of our toy model against predictions for the classical case ($j_d\rightarrow j_t$). In the top panel we show the number of steps needed to reach convergence (with a minimum of 5000, and maximum of 50,000). Unconverged simulations are shown as small black dots in later panels. In the second panel we show the fraction of systems which disrupt each step, $\Gamma_{cl}$. The final two panels show this fraction compared to the expectation from classical loss cone theory in the full regime (third panel) and empty regime (fourth panel). The dashed diagonal line shows $\sigma=j_t$, corresponding to $q\sim 1$, the approximate transition between the two regimes.}
    \label{fig:classic_rates}
\end{figure}

In figure \ref{fig:classic_rates} we show the equilibrium state for an ensemble of toy models of various $j_t$ and $\sigma$. We start from the approximate equilibrium solution presented in the last section, and evolve for a minimum of 5000 steps, until they either reach a steady state equilibrium or a maximum step number of 500,000. 

Examining the first panel, showing the number of steps needed to reach equilibrium, we see that most simulations reach equilibrium at, or before, the minimum number of steps. However for small $\sigma$ and/or $j_t$ the number of steps grows rapidly, and a small minority do not reach equilibrium within 500,000 steps.

The second panel shows the simulated $\Gamma_{cl}$ for all simulations that converge, and the latter two panels show $\Gamma_{cl}$ normalised by our expectation for the full and empty regime respectively.

There is a geometric factor introduced to the rate in the full regime (equation \ref{eq:Gamma_cl_full}) that depends on the shape of the physical region and the choice of $\Delta$. This is easiest to calculate in the case of a top-hat $\Delta$ function of width $2\alpha$. In this case is a region of $j_z,j$ space of that occupies fractional area $\alpha^2$ for which the probability of being scattered into the loss cone is $j_t^2/(4\alpha^2)$, and hence $\Gamma_{cl} \sim \frac{1}{4} j_t^2$. More complex $\Delta$ requires performing this integral by hand, and hence it is this simple case that we show in figure \ref{fig:classic_rates}. For $\sigma \ll j_t$ most walkers do not see the geometry of any boundaries, and there is no equivalent geometric correction in the empty regime.

We see that the prediction for $\Gamma_{cl}$ agree well with the numerical results in both the full and empty regimes (above and below the $q\sim 1$ line respectively). There are two regions showing a slight divergence in the empty regime for small $\sigma$. An enhanced $\Gamma_{cl}$ is seen at very large $j_t$, where geometric effects can once again become important. For small $j_t$ the predicted value is slightly higher than observed, this may be a sign that the simulation has not yet truly converged. 

It is interesting to note that, because we work with $p$ just after loss, this is equivalent to the moment just after periapse, when the loss cone is always empty. This initially seems to contradict the typical full loss cone, where there can be a significant transient population of systems with $j<j_d$ at any random time. However, as we have shown we recover the full loss cone behaviour, and indeed if we were to examine $p^S$ or $p^D$, we would see non-zero occupation of C.

Overall the agreement is strong, giving us confidence that this toy model reproduces the broad features of loss cone dynamics, and that extending it to $j_d>j_t$ will be illustrative of the true combined behaviour of scattering and diving orbits.


\subsection{Combined formalism}

For $j_d >j_t$ the net flux of walkers through the boundary of the diving region boosts the rate of disruptions.

To quantify this, we find the equilibrium state for a broad range of $j_t$, $j_d$ and $\sigma$ (using a Gaussian scattering kernel) giving an equilibrium disruption fraction $\Gamma(j_t,j_d,\sigma)$. We sample uniformly in log space for $10^{-2}<j_d<0.75$, $10^{-2}<\ln(j_d/j_t)<10$ and $10^{-3}<\sigma<10^{-1}$. For every simulation we also simulate the equivalent classical case giving $\Gamma_{cl}(j_t,\sigma)$ for comparison.

\subsubsection{Comparing to $\gamma_{\rm max}$}

When formulating our approximate solution we find a maximum curvature, $\gamma_{\rm max}(j_d)$, such that $p' \geq 0$ everywhere. We can compare this to the value of $\gamma$ obtained from $\Gamma$ for the numerical equilibrium solutions. 

\begin{figure}
\includegraphics[width=0.95\columnwidth]{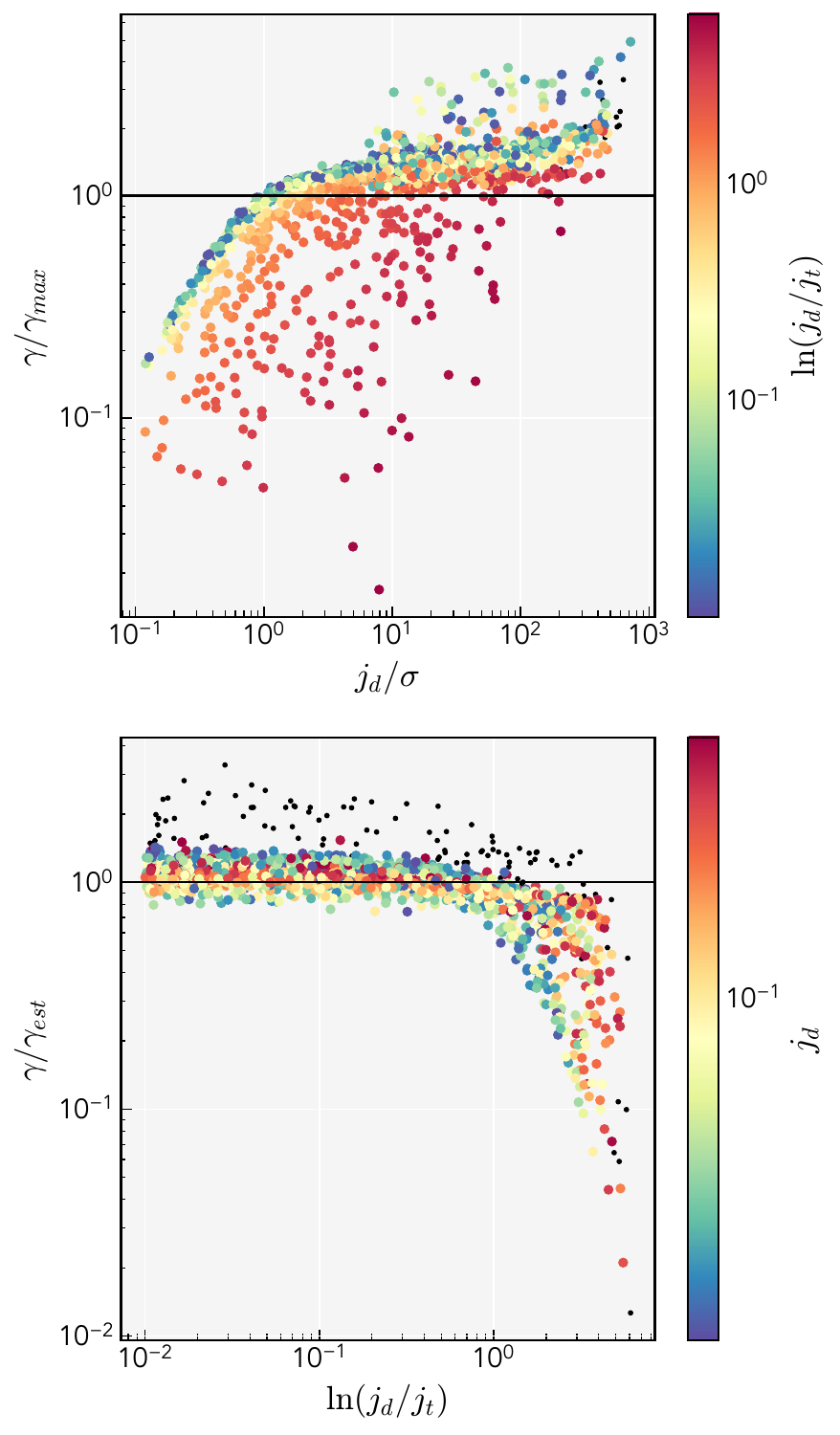}\\
    \caption{Comparison of the curvature parameter derived from our numerical equilibrium solutions, $\gamma = \sqrt{2\Gamma}/\sigma$, with the maximum curvature of our approximate solution ($\gamma_{max}$, equation \ref{eq:gamma_max}, top panel) and the improved estimate taking into account full and empty loss wedge behaviour ($\gamma_{est}$, equation \ref{eq:gamma_est}, bottom panel). A small number of simulations do not reach equilibrium within the maximum number of steps, shown as black dots in the first panel. In the second panel we also exclude simulations with $j_d > 0.575$, where the geometry of N begins to deviate significantly, which correspond to almost all the black points with $\gamma$ significantly greater than $\gamma_{est}$.}
    \label{fig:gamma_comp}
\end{figure}

We show this comparison in the first panel of figure \ref{fig:gamma_comp}, where we see that $\gamma_{max}$ is a relatively good estimator of the eventual rate, mostly well within an order of magnitude, though it diverges for small $j_d/\sigma$ and large $\ln(j_d/j_t)$. There is also a collection of simulations with poor agreement with $j_d \gtrsim 0.575$ (the maximum curvature solution with positive gradient), suggesting that some of our geometric arguments may break down for $1-j_d \ll 1$, where the N region goes to a narrow horizontal strip.

The behaviour with $j_d/\sigma$ suggests a transition analogous to the full and empty loss cone in the classical case. For $\sigma \ll j_d$ we are diffusion limited and thus have an \textit{empty loss wedge} with high curvature, and for $\sigma \gg j_d$ we have a \textit{full loss wedge} with low curvature. Examining the behaviour of $\gamma/\gamma_{max}$ we can posit a better estimate:
\begin{equation}
\label{eq:gamma_est}
\gamma_{est}(j_d,\sigma) = \frac{3}{2}\frac{\gamma_{max}}{1+\frac{\sigma}{j_d}}
\end{equation}
which we show in the second panel agrees excellently for all $\ln(j_d/j_t) \lesssim 1$, and overestimates the rate for $\ln(j_d/j_t) \gtrsim 1$. 

This estimate depends only on $j_d$ and $\sigma$. Its accuracy depends on being in the regime where $j_d/j_t$ is close to 1, but the fact that it is otherwise independent of $j_t$ shows that in this regime the diving region behaves like a slight enlargement of the loss cone. The factor of $\sigma/j_d$ captures the depletion of the area close to the diving region in the diffusion limited case, analogous to the factor of  approximately $|\ln j_t|/q$ in the classical loss cone formalism. In contrast here the diffusion at the edge of the diving region is essentially 1 dimensional, and thus the depletion factor scales linearly rather than logarithmically.

The factor of $3/2$ is chosen empirically, such as to bring the estimate into agreement with the measured rates, but likely represents the fact that our approximate solution does not have a self-consistent boundary layer. Such a boundary layer will primarily increase the gradient, $\partial_j p$, as $j\rightarrow j_d$, and thus we would expect a larger flux into the diving region, and thus a higher refill rate and larger curvature, in line with our pragmatic adjustment factor.

As $\ln(j_d/j_t)\rightarrow 0$ the diving region becomes a thin layer above the loss cone, and any solution should asymptote to the classical case. Thus the good agreement here shows that our approximate solution, with $p(0,j_t)\rightarrow 0$, captures the classical case very well.

A smaller $\gamma$ will give $p(0,j_t) >0$ and thus we can surmise that as the diving region becomes large there is a significant probability density along the boundary of the loss wedge, requiring a more complex solution that we have only been able to obtain numerically.

It is interesting to note that the limit on $\gamma_{max}$ is not directly an argument about the rate of disruptions, but the curvature of the solution. We could also calculate a $\Gamma$ for our approximate solution (see for example the initial $\Gamma$ in figure \ref{fig:p_Gamma_N}), but we find that this is a much less precise estimate of the true $\gamma$.

\subsubsection{Comparing to $\Gamma_{cl}$}

Another benchmark to compare to is how much the inclusion of the diving region boosts the rate of disruptions compared to the classical case.

\begin{figure}
\includegraphics[width=0.95\columnwidth]{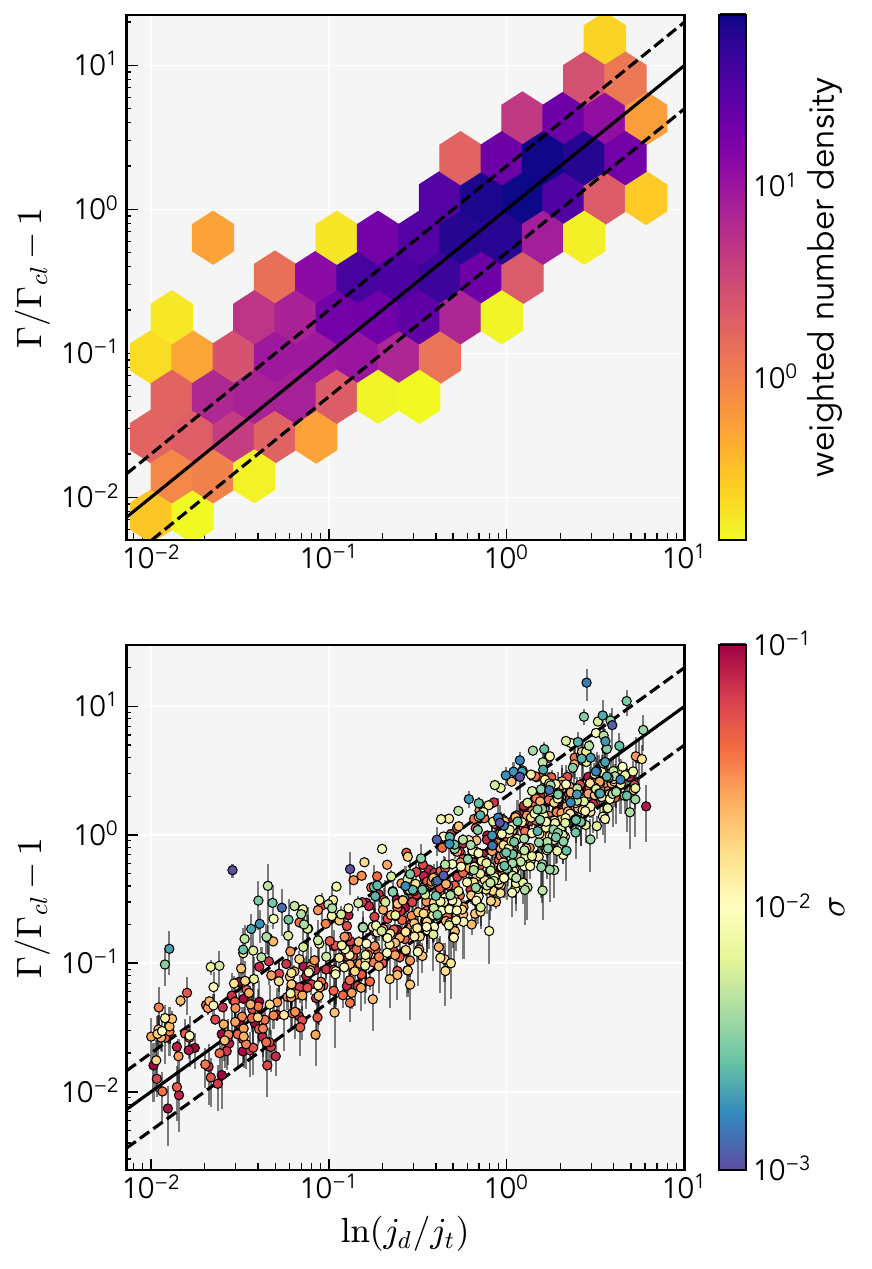}\\
    \caption{Boost to the fraction of disrupted systems per orbit, $\Gamma$, compared to the classical case, $\Gamma_{cl}$, for a wide range of $j_t$, $j_d$ and $\sigma$. We find that the boost is well approximated by the factor $\ln(j_d/j_t)$ and hence plot the approximately linear relationship between these (solid black line). Over and underestimates of a factor of $\frac{1}{2}$ and $2$ are shown as dashed black lines. In the top panel we show the density in this space, weighted by signal to noise, and in the bottom we show each individual simulation coloured by $\sigma$.}
    \label{fig:boost_results_rel}
\end{figure}

We find the useful empirical relationship, that the relative boost to the fraction of disruptions is well-approximated by a Coulomb-logarithm-like factor:
\begin{equation}
\label{eq:boost}
\frac{\Gamma}{\Gamma_{cl}} \approx 1+\ln\left(\frac{j_d}{j_t}\right).
\end{equation}
Figure \ref{fig:boost_results_rel} shows that this heuristic is obeyed over orders of magnitude of $\ln(j_t/j_d)$ and seemingly holds for all $\sigma$, though the error and scatter is larger for small $\sigma$.

Unlike $\gamma_{est}$ this empirical relationship holds for all $\ln(j_d/j_t)$, and immediately shows the magnitude of the impact of including (or ignoring) diving orbits. We've shown that $j_d$ can commonly reach values approaching unity, whilst $j_t$ is expected to be orders of magnitude smaller. For $j_t =10^{-3} j_d$, $\Gamma \approx 8 \Gamma_{cl}$ and for $j_t =10^{-6} j_d$, $\Gamma \approx 15 \Gamma_{cl}$. This is a significant boost, though the logarithmic behaviour makes the boost surprisingly insensitive to the order of magnitude of $j_t$, with $\Gamma \sim 10 \Gamma_{cl}$ being a relatively robust boost for any tidal disruption event of a trajectory with $j_d \gtrsim 0.1$. 

\subsubsection{Comparing to full and empty loss cone behaviour}

\begin{figure}
\includegraphics[width=0.95\columnwidth]{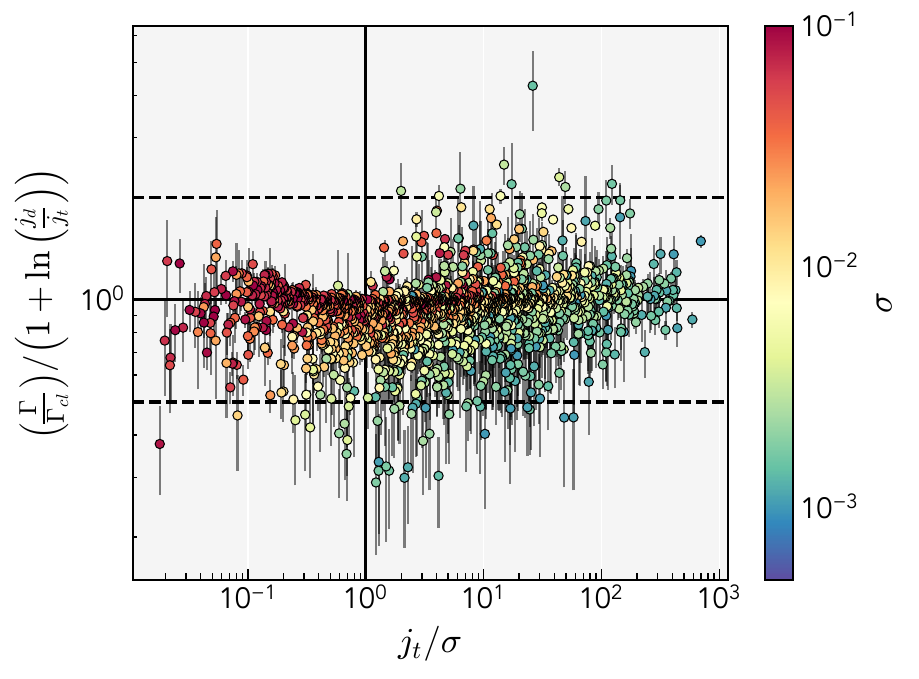}\\
    \caption{Comparing the empirical boost presented in figure \ref{fig:boost_results_rel} to $j_t/\sigma$. $j_t/\sigma =1$ (black line) approximately captures the transition between a full loss cone (left) and empty (right). A solid horizontal line shows perfect agreement between the predicted boost and that derived from the numerical equilibrium solution, and the dashed horizontal line show deviations by a factor of two above and below.}
    \label{fig:boost_empty_full}
\end{figure}

In figure \ref{fig:boost_empty_full} we show how the agreement with the empirical boost factor from equation \ref{eq:boost} depends on whether we are in the full or empty loss cone regimes. Generally the heuristic agrees very well, within a factor of two for all but the smallest $\sigma$ simulations (where we would expect the largest numerical noise on a converged solution). It's interesting to see that in the full loss cone regime over estimates are rare, suggesting that there may be a extra dependence here currently unmodelled - however at all $j_t/\sigma$ the modal value is very close to 1, making our estimate a robust predictor regardless of the classical loss cone regime.

\subsubsection{Comparing fluxes into the loss cone}

Finally it is interesting to ask whether the flux into the loss cone is dominated by diffusion through D from large $j_z$, or straight into W from small $j_z$. To quantify this, we can calculate the net flux from D to W, $\bar{\Gamma}_{DW}$ and compare this to the total flux into the loss cone. As shown in equation \ref{eq:Gamma_W} the total flux into W, and eventually into C, should be dominated by $\bar{\Gamma}_{DW}$ and $\bar{\Gamma}_{NW}$. Thus when $\bar{\Gamma}_{DW}/\Gamma$ is small we can conclude that flux directly into the wedge dominates, whilst when this ratio is large diffusion through D dominates. 

This ratio has direct implications for the properties of any disruption event and the associated observables/remnants. If diffusion through D dominates we can expect most disruptions to have $j_z$ preferentially close to $j_t$. Meanwhile if flux directly from N to W dominates we can expect a range of $j_z <j_t$ for disruptions, including (in some cases even dominated by) $j_z$ close to 0. As detailed in PRS25 section the distribution of $j_z$ at disruption directly informs the range of possible orientations and depths of disruption events.

\begin{figure}
\includegraphics[width=0.95\columnwidth]{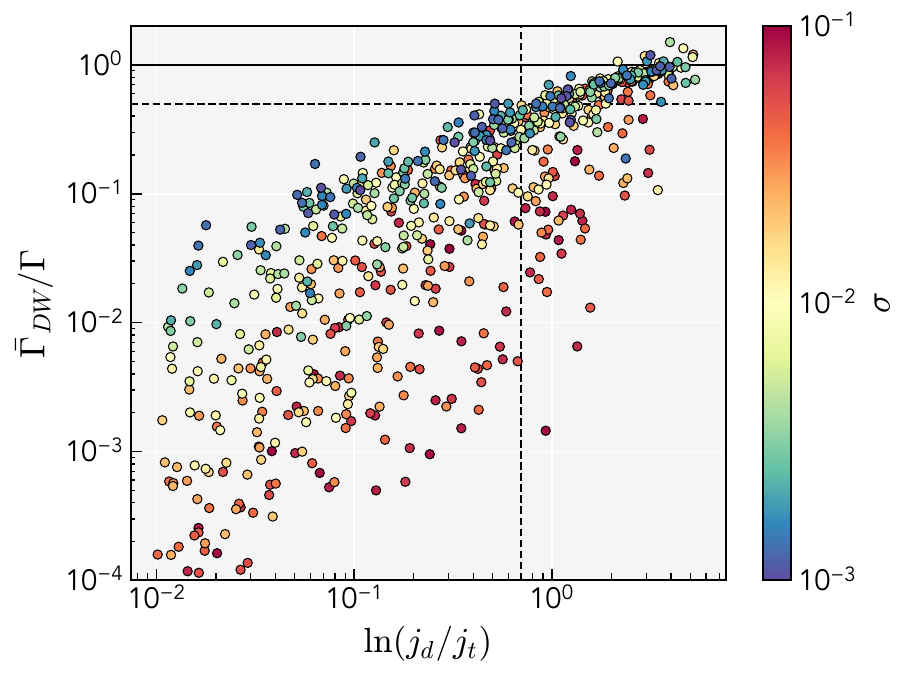}\\
    \caption{The fraction of the total flux into the loss cone which passes through the boundary between regions D and W. The majority of the remaining flux flows directly from N to W. The point at which the net flow can be dominated by flux from D to W (dashed horizontal line) corresponds to the point where $j_d \geq 2 j_t$ (dashed vertical line). The solid horizontal line shows where flux from D to W completely dominates (though some data points fall above this line due to Poisson noise).}
    \label{fig:flux_ratio}
\end{figure}

In figure \ref{fig:flux_ratio} we show the variation of this ratio as a function of $\ln(j_d/j_t)$. For small to moderate $\sigma$ this ratio scales close to linearly, which matches the intuitive result that the relative flux ratios should roughly scale with the length of the associated boundary. Accordingly the first solutions to be dominated by flux from D to W are those with $j_d \gtrsim 2 j_t$. For large $\sigma$ this geometric intuition breaks down, with a higher fraction of the flux coming directly from N to W. Generally speaking we would expect systems with $j_d \lesssim \sigma$ to have diminished flux directly from D to W.

It is important to note that even when $\bar{\Gamma}_{DW}$ is relatively small this does not mean that the diving region does not have a significant effect on the rate of disruptions. Walkers can advect significantly faster through this region and thus lead to larger fluxes towards $j_z=0$ even if they are commonly scattered back out of D before they cross into W.

\subsection{Distribution of disruption properties}

We can characterise a disruption by its severity and orientation. The severity is measured by the diving factor,
\begin{equation}
\beta = \frac{r_t}{r_p} = \left(\frac{j_t}{j}\right)^2,
\end{equation}
which is large for very radial orbits that deposit the system deep into the tidal radius, and for small values ($\lesssim 1$) we have a grazing encounter that is unlikely to cause a full disruption.

The orientation is mostly encoded in the inclination of the orbit at the time of disruption,
\begin{equation}
i = \arccos\left(\frac{j_z}{j}\right),
\end{equation}
where $i=0$ $(=\pi)$ is an encounter aligned (anti-aligned) with the disk plane and $i=\frac{\pi}{2}$ is perpendicular to the disk.

The uniform occupation of $l$ for diving orbits means that $p(j)$ is constant, and allows us to directly convert these to probability distributions:
\begin{equation}
p(\beta|j_z) \propto p(j) \frac{dj}{d\beta} \propto j^{3} \propto \beta^{-\frac{3}{2}}.
\end{equation}
and 
\begin{equation}
p(\cos i|j_z) \propto p(j) \frac{dj}{d\cos i} \propto j^{2} \propto \left(\cos i \right)^{-2}.
\end{equation}
Normalizing these distributions given their limiting values gives
\begin{equation}
p(\beta|j_z) = \frac{j_t}{2(j_t-j_z)} \beta^{-\frac{3}{2}} \mathrm{\quad for \quad} 1<\beta<\left(\frac{j_z}{j_t}\right)^{-2}
\end{equation}
and
\begin{equation}
p(\cos i|j_z) = \frac{j_z}{2(j_t-j_z)} \left(\cos i \right)^{-2} \mathrm{\quad for \quad} \frac{j_z}{j_t}<|\cos i| <1
\end{equation}
(these were originally derived in PRS25 section 5, though are more concisely expressed here using the uniform occupation of $j$ space).

Now that we have found numerically the form of $p(j_z)\propto(j_t-j_z)\rho(j_z)$ we can marginalize this over $j_z$ to give
\begin{equation}
p(\beta) = \int p(\beta|j_z) p(j_z) dj_z \propto \int_0^{j_t \beta^{-\frac{1}{2}}} \rho(j_z) \beta^{-\frac{3}{2}} dj_z 
\end{equation}
and
\begin{equation}
p(\cos i) = \int p(\cos i|j_z) p(j_z) dj_z \propto \int_0^{j_t} j_z \rho(j_z) \left(\cos i\right)^{-2} dj_z. 
\end{equation}

In one extreme case, of a completely flat $\rho(j_z) = const.$ we recover $p(\beta)\propto \beta^{-2}$ and isotropic disruptions, $p(\cos i) = \rm const$. In the other extreme case, where the only systems entering the loss cone are doing so at the outer edge of the loss wedge, $p(j_z) \rightarrow \delta(j_z-j_t)$, we find $p(\beta) = \delta(\beta-1)$ and $p(|\cos i|) = \delta(|\cos i| -1)$ (i.e. all disruptions occur exactly in the disk plane aligned or anti-aligned with the potential). This is an analogous dichotomy to the empty and full loss cone in the classical case.

In our framework we expect $\rho(j_z)$ to be a monotonically increasing function of $j_z$ (as net flow is always towards lower $j_z$ and disruption) and thus there is no situation in which we see preferentially misaligned disruptions ($\cos i \sim 0$). This is in contrast with PRS25, where we did not fully model the depletion of $j_z$, and argued that the more efficient disruption of low $j_z$ orbits would lead to a strong excess with $j_z \sim 0$.

\begin{figure*}
\includegraphics[width=0.95\textwidth]{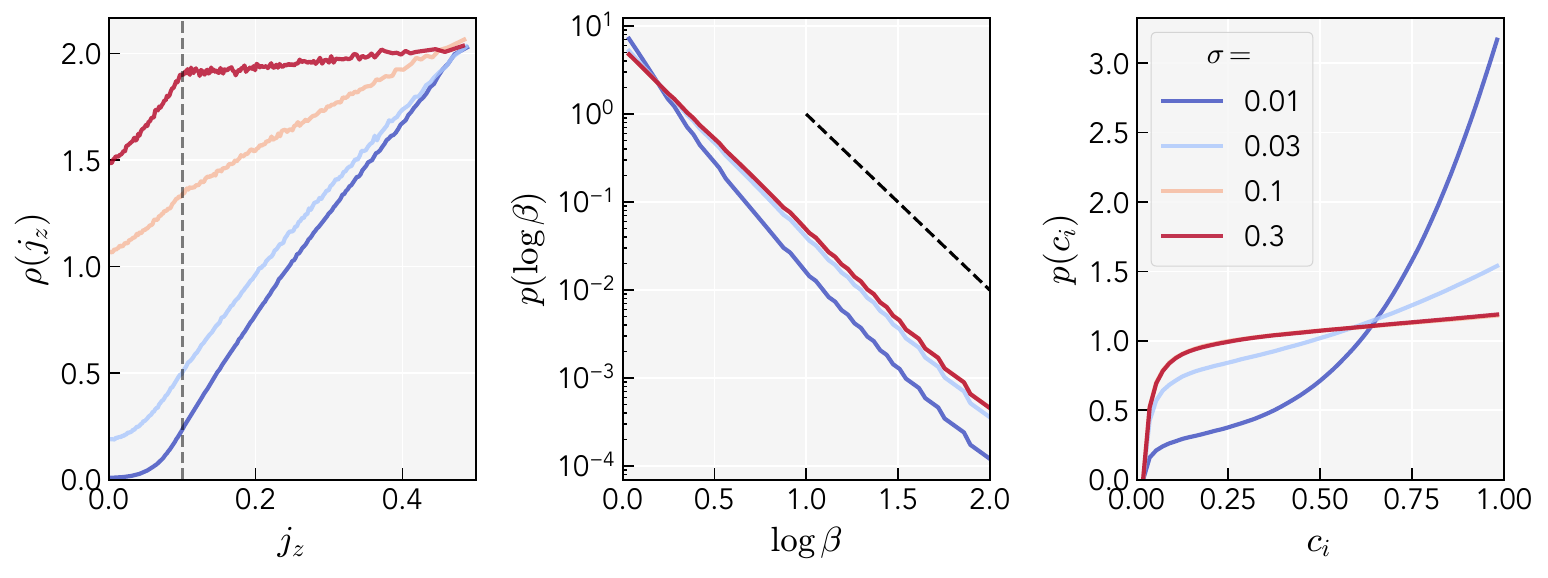}\\
    \caption{The probability distribution, $\rho(j_z) = p(j_z,j)$ within the diving region, and the associated distribution of disruption properties: diving factor, $\beta$, and inclination, $i$. We hold $j_t=0.1$ and $j_d=0.5$ constant and show the equilibrium state for 4 different $\sigma$ values. The dashed black line in the $\beta$ distribution shows a slope of $p(\beta)\propto \beta^{-2}$.}
    \label{fig:rho_dist}
\end{figure*}

In figure \ref{fig:rho_dist} we show some example equilibrium distributions of $\rho(j_z)$ and the corresponding distributions $p(\beta)$ and $p(|\cos i|)$. We see that when $\sigma$ is large the distribution of $j_z$ is almost flat, but drops off more steeply (and approximately linearly) in the loss wedge. For smaller $\sigma$ we see a steeper function, with more curvature in the loss cone. It is only for the smallest $\sigma$ shown that we see a vanishing $\rho(0)$.

The distribution of $\beta$ only differs perceptibly from the $p(\beta)\propto \beta^{-2}$ distribution for the most depleted case. The distribution of $|\cos i|$ is more sensitive, with (anti)aligned disruptions significantly preferred for the two smallest $\sigma$ cases. Even the slight in the loss wedge seen for the largest $\sigma$ is sufficient to discourage perfectly misaligned orbits, but otherwise the distribution tends to isotropy for large $\sigma$.

\begin{figure}
\includegraphics[width=0.95\columnwidth]{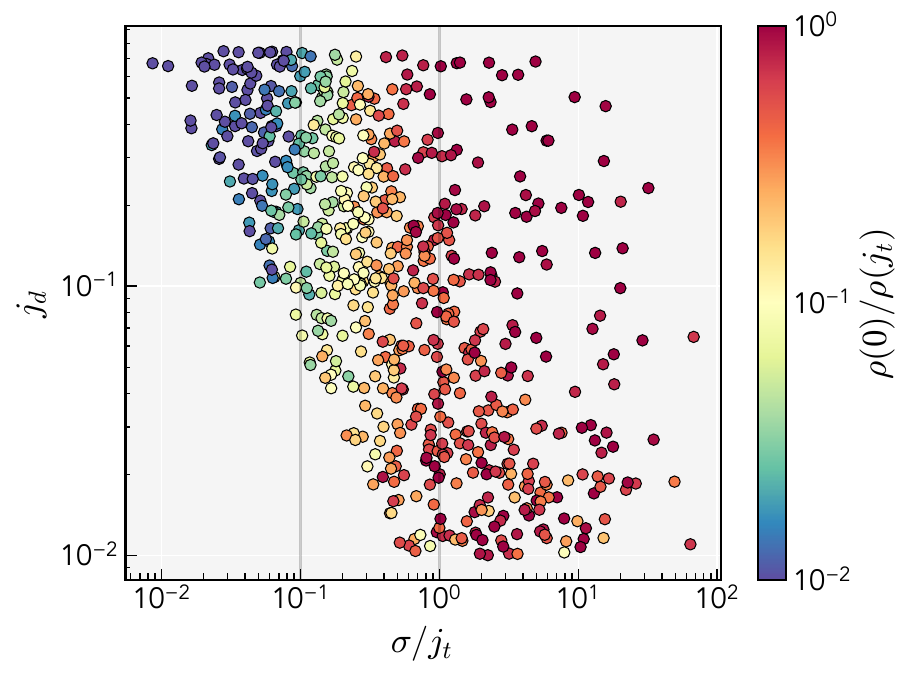}\\
    \caption{The ratio of $\rho(j_z)$ between the edge of the loss wedge ($j_z=j_t$) and the center ($j_z=0$), which captures how depleted the inner loss wedge is. Vertical lines at $\sigma/j_t=1$ and $0.1$ denote the approximate regions outside of which there is $\sim$ no/full depletion.}
    \label{fig:rho_ratio}
\end{figure}

We can more precisely quantify the transition to depletion of the inner loss wedge by comparing $\rho$ at $j_z=0$ to $j_z=j_t$ over a wide range of parameters, as shown in figure \ref{fig:rho_ratio}. Here we can see near total depletion for $\sigma \lesssim 0.1 j_t$, and no depletion for $\sigma \gtrsim j_t$. The dependence on $j_d$ is very slight (though the overall normalization of $\rho$ and subsequent $\Gamma$ are more strongly dependent) but systems with larger $j_d$ are marginally more depleted.

We note that the probability distributions here are given for a fixed energy, and to recover the full distributions and total rates of disruption we would expect for a galaxy we would need to know $p(\epsilon)$ and marginalize also over energy.

\section{Discussion and Conclusions}
\label{sec:discuss}

In this paper we present two major conclusions
\begin{itemize}
\item Chaotic diving orbits are ubiquitous at low angular momentum for axisymmetric potentials with even relatively moderate flattening, affecting most families of trajectories that traverse the transition between regions dominated by spherical and axisymmetric potentials. We can approximate the effect of these orbits into a single $j_d(\epsilon, l_z)$, the relative angular momentum below which we expect approximately all orbits to be chaotic and vary between $j_z<j<j_d$ orbit to orbit
\item Systems on these chaotic diving orbits can advect significantly faster towards low total angular momentum and eventual disruption, occurring when their total relative angular momentum $j$ goes beneath a critical value $j_t$ at periapse. We present analytical arguments and toy models that estimate the boost to disruption rates, $\Gamma$, compared to the classical case which ignores diving orbits, $\Gamma_{cl}$, of approximately $\Gamma/\Gamma_{cl} \sim 1 +\ln(j_d/j_t)$, a factor that can increase rates by an order of magnitude.
\end{itemize}

This work extends the conclusions of \citet{Penoyre25} (PRS 25), showing the existence of these chaotic diving orbits over a much broader range of potentials, and more precisely quantifying their extent and impact. We refine and improve the method for identifying the scope of chaotic orbits. We greatly increase the efficiency by only simulating those orbits which are likely to be chaotic and diving. We also markedly improve the accuracy of identification with a new technique for identifying whether the surface of section is 1D (regular) or 2D (chaotic) as detailed in appendix \ref{ap:dim}.

In PRS25 we performed a simple calculation for the relative boost to the rate of disruptions, finding an increase of one or two orders of magnitude overall depending on $j_t$. That estimate did not fully consistently take into account the depletion at the boundary of the diving region, and the mutual interplay of classical two body scattering events and chaotic redistribution of angular momentum. The boost, relative to the classical case, that we find here is more moderate though still large, and also more robustly estimated.

Unlike previous literature on the loss cone (e.g. \citealt{Cohn78}) we find a global (if simplified) equilibrium solution. We do this by modelling the probability of occupation of parameter space, which is always normalised to 1, rather than the number density, which decreases over time. In doing so we show that the renormalisation step, accounting for disruption, behaves like a positive flux proportional to the local probability density, and thus our solutions depend on a curvature parameter that links the rate of disruptions at very small $j$ to the Laplacian across all of the parameter space. This removes any time dependence from the solution (though a consistent loss rate and diminishing number of systems can be calculated) and may more closely represent the current state of any galaxy which has been able to evolve slowly enough to remain close to equilibrium. We also show that all of our results, where applicable, are compatible with the classical loss cone formalism, recovering the expected rates and, if we ignore refilling, number density distribution.

We calculate the distribution of disruption outcomes, parametrised by how deep the system dives into the tidal radius, $\beta$, and the inclination relative to the galactic plane, $i$. When there is relatively uniform occupation throughout the loss wedge disruptions, which occurs when $\sigma \gtrsim j_t$, we see close to isotropic disruptions with a slight depletion at $i\sim 0$, and $p(\beta)\propto \beta^{-2}$. When the inner loss cone (near $j_z=0$) becomes very depleted, which occurs when $\sigma \lesssim 0.1 j_t$, disruption more frequently have low $\beta$ and inclinations close to perfectly aligned or anti-aligned with the potential (i.e. in the disk plane).

Our numerical estimates span a wide range of physical regimes, and the boost to disruption rates is robust regardless of whether we are in the classical full or empty loss cone regimes. We are also able to identify analogous full and empty loss wedge regimes, depending on whether the orbit averaged diffusion in angular momentum is small or large compared to $j_d$. As $j_d$ can be of order unity, whilst the rate of diffusion and $j_t$ are typically very small (the latter scaling as the square root of the ratio of tidal radius to orbital radius, AU compared to kpc) we expect most physical disruptions be in the diffusion limited empty loss wedge regime.

An increase of this magnitude to the predicted rate of tidal disruptions may either be a boon or a challenge to the community. If the rate of disruptions is known empirically then it can be matched by diminishing the expected population of disrupting objects, for example reducing our expectation for the number of candidate binaries in the galactic center that can form HVSs. 

In other cases an increased predicted rate may be in direct tension with observations. The current predicted rate of TDEs, before including the boost detailed in this paper, is in possible tension with observations \citep{Stone16}, a problem only exacerbated by consistently including the effect of diving orbits. Partial disruption events may provide a mollification, corresponding to more frequent moderately close passages to the MBH, which tidally distort but do not necessarily completely disrupt systems. Partial tidal disruptions of stars (e.g. \citealt{Guillochon14}) may strip small amounts of mass over a series of passages, rather than in a single cataclysmic event, and thus not liberate enough energy quickly enough to be visible extragalactically. Mass and energy loss can also change the orbit, and thus lead to less severe future close passages, as is shown to be the case for 'fly-away' binaries \citep{Sersante25} which are forced closer together by the interaction with the MBH, liberating enough orbital energy that they do not return for subsequent close passages.

The degree of flattening of the central potential of a galaxy links directly to the rate and character of diving orbits. Thus in galaxies where diving orbits provide a significant, even dominant, boost we may be able to infer, or at least constrain, the shape of the potential inner regions of distant galaxies.  

In PRS25 we used a slightly extended potential for the central MBH (necessitated by working with models that inferred potentials based on the density profile). In this paper we use a true point mass MBH and find that whilst chaotic diving orbits extend into the sphere of influence they are strongly suppressed and eventually disappear. The innermost chaotic orbits have $\rho \sim 1/2$, corresponding to an apoapse just outside of the sphere of influence. This means that the extent of the sphere of influence, and the relative dominance of the MBH over other object in the GC, directly imprints itself on disruption rates. If the mass of a MBH can be well estimated we may be able to constrain the mass of other stars, compact objects and dark matter in its close vicinity, differentiating central density cores and cusps.

Though chaotic orbits are suppressed very close to the MBH there is still the capacity for orbits to precess into the loss cone due to the small perturbative contribution of the flattened external potential \citep{Kaur25}. Such orbits could be thought of as non-chaotic diving orbits, and extending the paradigm to within the sphere of influence will be important to calculate consistent total disruption rates. There is a practical lower limit to the degree of chaotic behaviour that can be detected using the framework presented here, and thus it is also possible that a marginal population of chaotic orbits may extend further into the sphere of influence.

The boost factor we calculate in this paper is specific to each energy and z-angular momentum, and to find a total rate for any given galaxy or species of object requires integrating this over the demographics of a population. By concisely summarising the effect of chaotic diving orbits we aid that process, but we save enacting it for future work and other readers.

\section*{Data Availability}
The data underlying this article will be shared on reasonable request to the corresponding author.

\section*{Acknowledgements}
We would like to thank Emily Sandford for sharing expertise on chaotic attractor reconstruction, Yihan Wang for performing corroborating high-precision orbit integrations, Mark Dodici for discussions about disruption outcomes, and Mara Boers and Elise Yu for helping to clarify the notation used for $p_\theta$. zp and EMR acknowledge support from European Research Council (ERC) grant number: 101002511/project acronym: VEGA P.

\bibliographystyle{mnras}
\bibliography{bib}
\bsp

\appendix

\section{Dimensionality as a test of chaotic behaviour}
\label{ap:dim}

Following e.g. \cite{grassberger1983, grassberger1985}, we can define the correlation dimension, $D$, of a distribution of objects, $n(\mathbf{r})$ based on how the number of objects, $N(<r)$, within some M-sphere of radius $r$, varies as a function of $r$:
\begin{equation}
\begin{aligned}
N(<r) &= \int n(\mathbf{r}) d^M\mathbf{r}= \int_0^r \frac{M\pi^\frac{M}{2}}{\Gamma_E(\frac{M}{2}-1)} r^{M-1} n(\mathbf{r}) d r \\
&\propto r^{\max(D,M)} 
\end{aligned}
\end{equation}
\label{lastpage}
where $\Gamma_E$ is Euler's Gamma function. 

To give an example, close to the sun the number of stars within a 3D radius $r$ scales as roughly $N(<r)\propto r^3$, as the density within the galactic disk is relatively constant. For larger distance, beyond the scale height of the disk, we might expect $N(<r) \propto r^2$ (though this simple prediction is modified by the varying density with galactic radius and small but non negligible populations of halo stars outside of the disk).

For a finite region, $r \subset R$, and corresponding finite number of objects $N=N(<R)$ we can rewrite this in term of fractional distance
\begin{equation}
d=\frac{r}{R} \ (\leq 1)
\end{equation}
and
\begin{equation}
F=F(<d)=\frac{N(<d)}{N} \ (\leq 1)
\end{equation}
(equivalent to the cumulative distribution function) such that
\begin{equation}
\label{eq:ap_scaling}
F \sim d^D
\end{equation}
where we have assumed that we're working in a sufficiently high dimensional space that $M\geq D$.

Thus we can calculate $D$ from
\begin{equation}
\label{eq:ap_D}
D(d) =\frac{d \log F}{d\log d}.
\end{equation}
If there is a consistent dimensionality then this equation should hold over a wide range of scales, limited by Poisson noise at very small $d$, and the effect of the boundaries of the region as $d \rightarrow 1$.

This calculation is further complicated if the density profile is not uniform over the space it occupies, with centrally concentrated $n$ giving measured $D$ less than the true value.

\subsection{Application to Surfaces of Section}

In our 2D Surfaces of Section (SoS) we expect regular orbits to occupy (a family of) line segments with $D\rightarrow 1$. As we transform these regular orbits closer and closer to closed orbits, occupying single points, $D\rightarrow 0$. Chaotic orbits however are not constrained by a third integral of motion, and spread out over an area, with $1<D\leq  2$. Because our coordinate system is canonical, and we expect any chaotic trajectory to be ergodic, this area should be occupied uniformly.

Thus, for any given trajectory (integrated over many orbits) we should be able to choose some random point (away from any edge) and measure $F(<d)$ for a range of $d$ from this point and calculate $D(d)$ via equation \ref{eq:ap_D}. However this can break down if the occupied regions of the SoS are small and sparse, such that $F(<d)$ behaves almost like a step function, with $D(d)$ only taking meaningful values over some very small range of $d$. In theory if we repeated the process over many random initial points the average behaviour would converge to a universal $D$, however we can make this process much more efficient.

Instead we can calculate all point-to-point distances in ascending order, $d_i$, for each of the $K=k(k-1)/2$ pairs of our $k$ points in the SoS. Then we can calculate the analogous
\begin{equation}
\label{eq:ap_F}
F(d) = \frac{I}{K}
\end{equation}
where $I$ is the smallest $i$ such that $d_{I+1}>d$. This still satisfies equation \ref{eq:ap_D}, but preferentially samples distances that contain meaningful information about the behaviour of $D(d)$.

To assign a single $D$ for a given trajectory we need to choose some particular $d$ at which to measure it. We found it optimal to choose as small a value as possible whilst still large enough so as not to be dominated by Poisson noise. To capture this we choose the $100/\sqrt{k}$ percentile value of $d$. As numerical derivatives are highly susceptible to noise we smooth the gradient with a median filter of width 5.

\begin{figure*}
\includegraphics[width=\textwidth]{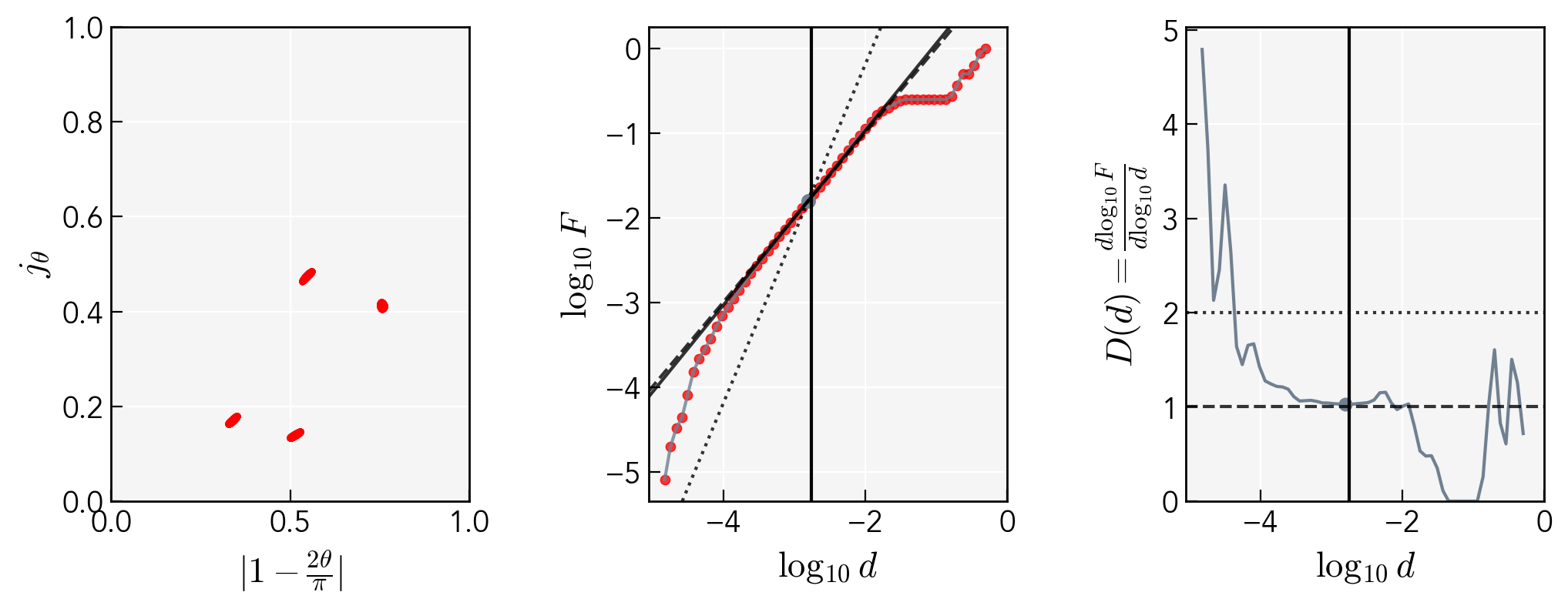}
    \caption{The dimensionality of orbit 8 from figure \ref{fig:map}. We show the SoS (left), the variation of $F(<d)$, as given by equation \ref{eq:ap_F}, with $d$ (middle), and the derived gradient (right). In the middle and right panels we show gradients corresponding to $D=1$ (dashed line) and $D=2$ (dotted line). The solid vertical line shows the value of $d$ where we measure $D$. The solid diagonal line in the middle figure shows the slope of the measured gradient at that point.}
    \label{fig:regular_dim}
\end{figure*}

\begin{figure*}
\includegraphics[width=\textwidth]{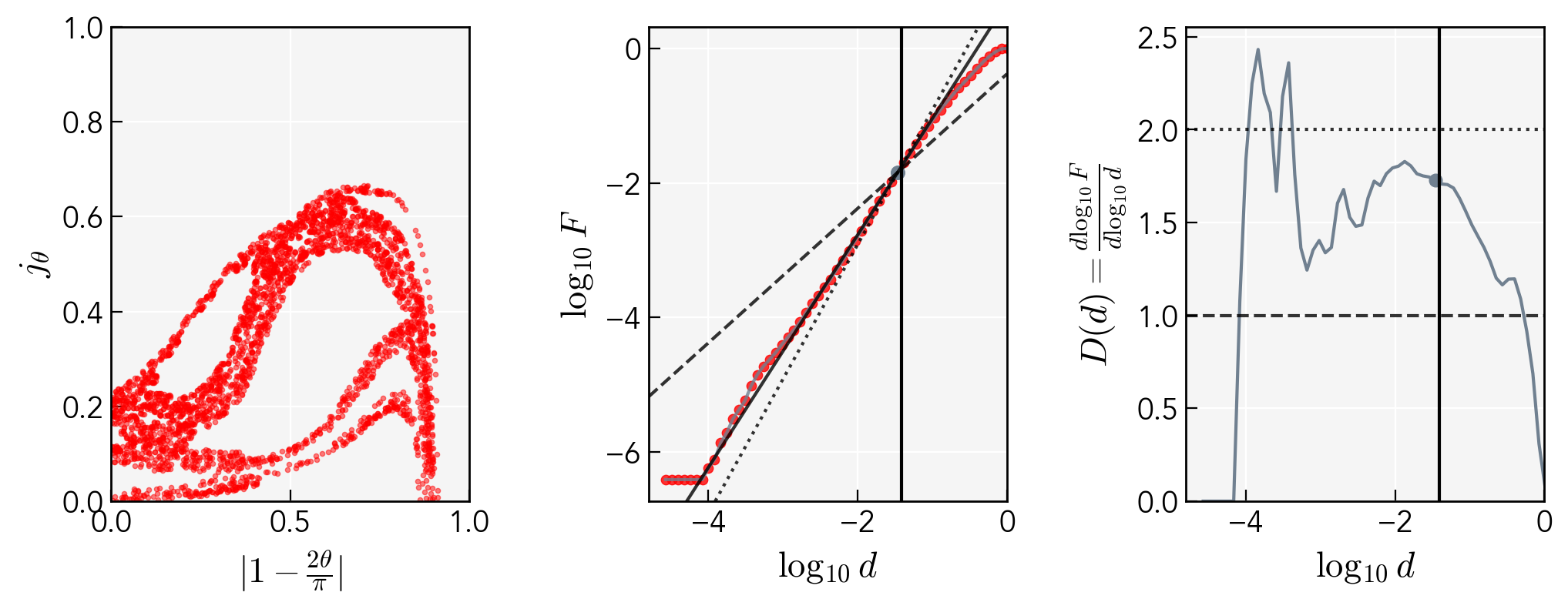}
    \caption{The equivalent of figure \ref{fig:regular_dim} but for a chaotic orbit (in the same potential).}
    \label{fig:chaos_dim}
\end{figure*}

Figures \ref{fig:regular_dim} and \ref{fig:chaos_dim} show the steps of this process for a regular and chaotic orbit respectively. The trajectory in figure \ref{fig:regular_dim} has 4 locii of points in the SoS, and examining the gradient we see that it is large at very low $d$ (Poisson noise dominated), but quickly tends to close to 1 over a large range of $d$ (including the value from which we derive $D$). At quite large d $(\sim 0.1)$ the gradient goes to zero, reflecting the fact that the small locii of points look like singular 0D points on scales larger than the locii but smaller than their separation.

The chaotic trajectory, figure \ref{fig:chaos_dim} consistently has a gradient steeper than 1 across almost all values of $d$. Examining the SoS we see that it is not perfectly uniform (either due to fractal substructure, or preferentially closely following a regular trajectory over many orbits) and thus we would expect $D<2$. We see that $D(d)$ peaks at about 1.5, and the measured value of $D$ is close to this value.

\begin{figure}
\includegraphics[width=\columnwidth]{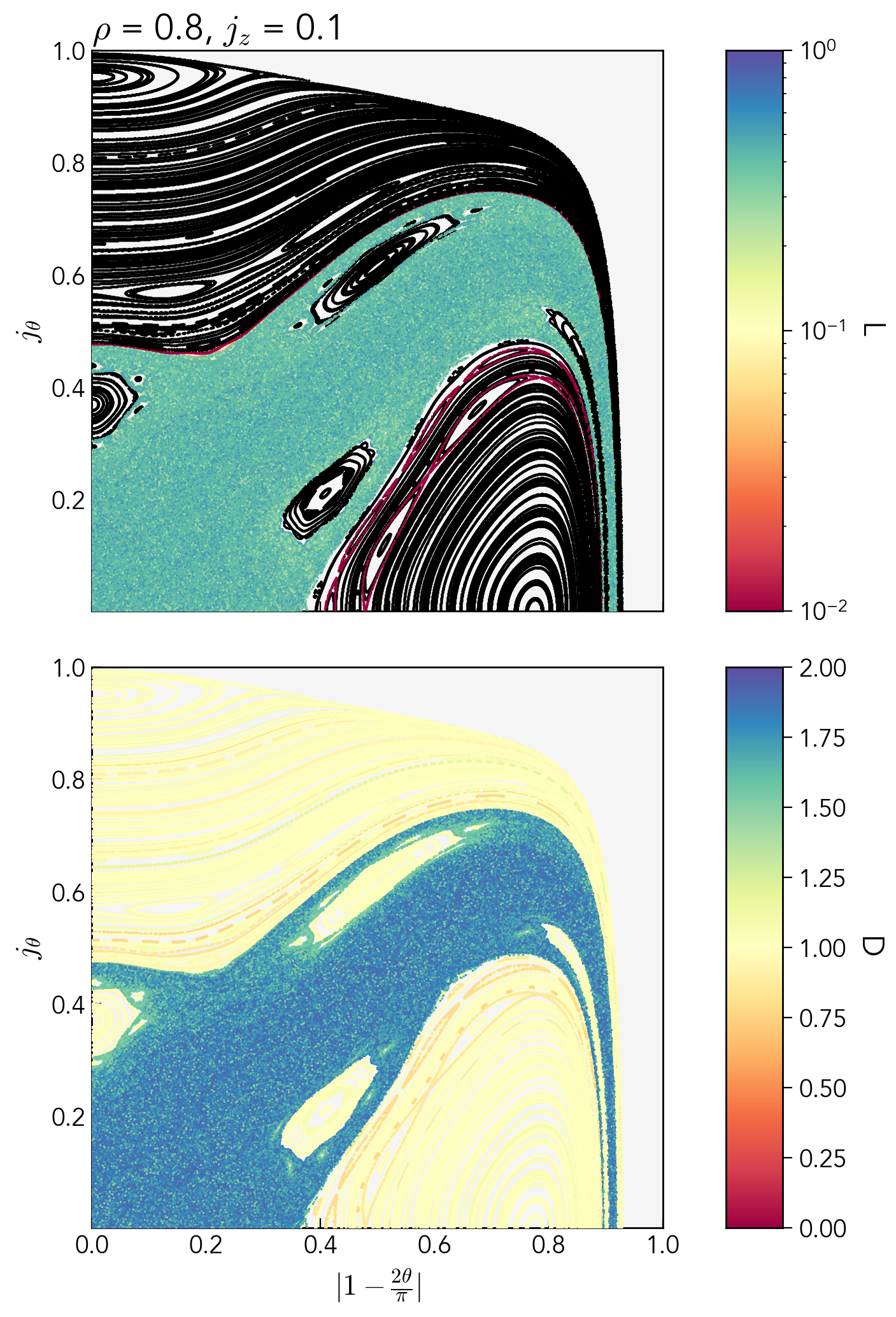}
    \caption{Comparing the Lyapunov exponent, $L$, as measured by \texttt{AGAMA} (upper panel) to our dimensionality metric, $D$ (lower panel). We use a model potential with $\mu=\frac{1}{30}$ and $\alpha=1$, and integrals of motion parametrised by $\rho=0.8$ and $j_z=0.1$. Trajectories are integrated for $\sim$2000 periapse passages. When measuring $L$ orbits classified as regular have no value of $L$, and are coloured black.}
    \label{fig:chaos_metrics}
\end{figure}

Finally in figure \ref{fig:chaos_metrics} we compare the two measures for a full family of trajectories. Almost all of the chaotic region is characterised by a high Lyapunov exponent, $L \sim 0.3$. This reduces for some trajectories which start (and remain) very close to the edges, where it may take many orbits to diverge from an almost regular trajectory. However we also see examples of regular orbits with very low but non-zero $L$($\lesssim 0.01$), especially close to harmonics of the banana orbit (bottom right of the SoS).

Generally we find regular orbit very reliably have $D \approx 1$ (or lower if we unluckily capture a length scale where the distribution appears to be 0D). Notably there are no misclassified regular orbits, and chaotic behaviour is detectable out to the edge of the chaotic region, though there is a slight reduction in $D$ ($\sim 1.5$). In general we find any value of $D>1.05$ to be a reliable flag of chaotic behaviour.

\end{document}